%% file: main.tex
\documentclass[acmsmall]{acmart}

\usepackage{algorithm}
\usepackage[noend]{algpseudocode}
\usepackage{framed}
\usepackage{subcaption}
\usepackage{xspace}
\usepackage{tikz}

\newcommand*\circled[1]{\tikz[baseline=(char.base)]{
    \node[shape=circle,draw,inner sep=0.2pt] (char) {#1};}}

\newcommand{\sysname}{Kepler\xspace}

\newcommand{\rewrite}[1]{\textcolor{black}{#1}}

\newcommand{\sm}{S^{\texttt{model}}}
\newcommand{\sopt}{S^{RCE}}

\newcommand{\freg}{P^{\texttt{reg}}}
\newcommand{\popt}{P_{99}^{RCE}}

\AtBeginDocument{%
  }

\setcopyright{rightsretained}
\acmJournal{PACMMOD}
\acmYear{2023} \acmVolume{1} \acmNumber{1} \acmArticle{109} \acmMonth{5} \acmPrice{}\acmDOI{10.1145/3588963}

\makeatletter
\gdef\@copyrightpermission{
  \begin{minipage}{0.2\columnwidth}
   \href{https://creativecommons.org/licenses/by/4.0/}{\includegraphics[width=0.90\textwidth]{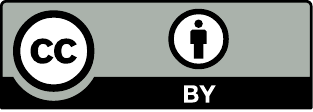}}
  \end{minipage}\hfill
  \begin{minipage}{0.8\columnwidth}
   \href{https://creativecommons.org/licenses/by/4.0/}{This work is licensed under a Creative Commons Attribution International 4.0 License.}
  \end{minipage}
  \vspace{5pt}
}
\makeatother

\begin{document}

\title{\sysname: Robust Learning for Faster Parametric Query Optimization}

\author{Lyric Doshi}
\authornote{Equal contribution.}
\email{lyric@google.com}
\orcid{0009-0002-1234-0283}
\affiliation{%
  \institution{Google}
  \city{Mountain View}
  \state{CA}
  \country{USA}
}

\author{Vincent Zhuang}
\authornotemark[1]
\email{vincentzhuang@google.com}
\orcid{0009-0007-2931-3069}
\affiliation{%
  \institution{Google}
  \city{Mountain View}
  \state{CA}
  \country{USA}
}

\author{Gaurav Jain}
\email{gaurav@gauravjain.org}
\orcid{0009-0006-9980-7737}
\affiliation{%
  \institution{Google}
  \city{Mountain View}
  \state{CA}
  \country{USA}
}

\author{Ryan Marcus}
\email{rcmarcus@seas.upenn.edu}
\orcid{0000-0002-1279-1124}
\affiliation{%
  \institution{University of Pennsylvania}
  \city{Philadelphia}
  \state{PA}
  \country{USA}
}

\author{Haoyu Huang}
\email{haoyuhuang@google.com}
\orcid{0000-0002-0940-228X}
\affiliation{%
  \institution{Google}
  \city{Mountain View}
  \state{CA}
  \country{USA}
}

\author{Deniz Altınbüken}
\email{denizalti@google.com}
\orcid{0000-0002-4558-2847}
\affiliation{%
  \institution{Google}
  \city{Mountain View}
  \state{CA}
  \country{USA}
}

\author{Eugene Brevdo}
\email{ebrevdo@google.com}
\orcid{0009-0005-7965-3534}
\affiliation{%
  \institution{Google}
  \city{Mountain View}
  \state{CA}
  \country{USA}
}

\author{Campbell Fraser}
\email{campbellf@google.com}
\orcid{0009-0005-2035-4409}
\affiliation{%
  \institution{Google}
  \city{Mountain View}
  \state{CA}
  \country{USA}
}

\renewcommand{\shortauthors}{Lyric Doshi et al.}

\begin{abstract}
\input{sections/0_abstract}
\end{abstract}




\maketitle

\input{sections/1_introduction}
\input{sections/2_related_work}
\input{sections/3_overview}
\input{sections/4_candidate_generation}
\input{sections/5_training_execution}

\input{sections/6_modeling}
\input{sections/7_experiments}

\input{sections/8_conclusion}

\received{July 2022}
\received[revised]{October 2022}
\received[accepted]{November 2022}

\bibliographystyle{ACM-Reference-Format}
\bibliography{main}

\end{document}

%% file: sections/0_abstract.tex
Most existing parametric query optimization (PQO) techniques rely on traditional query optimizer cost models, which are often inaccurate and result in suboptimal query performance. We propose \sysname, an end-to-end learning-based approach to PQO that demonstrates significant speedups in query latency over a traditional query optimizer. Central to our method is Row Count Evolution (RCE), a novel plan generation algorithm based on perturbations in the sub-plan cardinality space. While previous approaches require accurate cost models, we bypass this requirement by evaluating candidate plans via actual execution data and training an ML model to predict the fastest plan given parameter binding values. Our models leverage recent advances in neural network uncertainty in order to robustly predict faster plans while avoiding regressions in query performance. Experimentally, we show that \sysname achieves significant improvements in query runtime on multiple datasets on PostgreSQL. 

%% file: sections/1_introduction.tex
\section{Introduction}
Parametric query optimization (PQO) aims to optimize \emph{parameterized queries}, i.e. queries that have identical SQL structure and only differ in the value of bound parameters. Such parameterized queries are ubiquitous in modern database usage and present a significant opportunity for improving query performance because they are executed repeatedly. 

However, PQO has primarily been studied from the perspective of reducing query planning time by avoiding re-optimization when possible~\cite{alucc2012parametric, hulgeri2002parametric, ioannidis1997parametric, vaidya2021leveraging, chaudhuri2010variance, dutt2017leveraging}. Such approaches are implicitly constrained by the performance of the system's query optimizer, and therefore inherit all of the well-studied sub-optimalities of traditional query optimizers \cite{leis2015good}. Thus, an ideal system for parameterized queries should not only seek to minimize planning time via PQO, but also optimize query execution performance via query optimization (QO).

A variety of approaches have attempted to improve query optimization by applying machine learning \cite{yang2019deep, kipf2018learned, yang2020neurocard, negi2021flow, marcus2021bao}. Unfortunately, most learned query optimization techniques suffer from at least four drawbacks: (1) they require \emph{inference times} higher than traditional methods~\cite{lu2021pre, kim2022learned}, (2) they have \emph{inconsistent performance} across dataset sizes and distributions~\cite{kim2022learned, sun2021learned, marcus2019neo}, and (3) they often have \emph{unclear query performance improvements}~\cite{kim2022learned}. Worse yet, many of these learned systems lack (4) \emph{robustness}: regressions in query performance are unacceptable in most production scenarios~\cite{ding2019ai}. This poses an especially large challenge for learning-based approaches, since they typically cannot guarantee that all of their predictions result in improved execution time~\cite{wang2020we}.

\begin{figure}[t!]
    \centering
    \includegraphics[width=\linewidth]{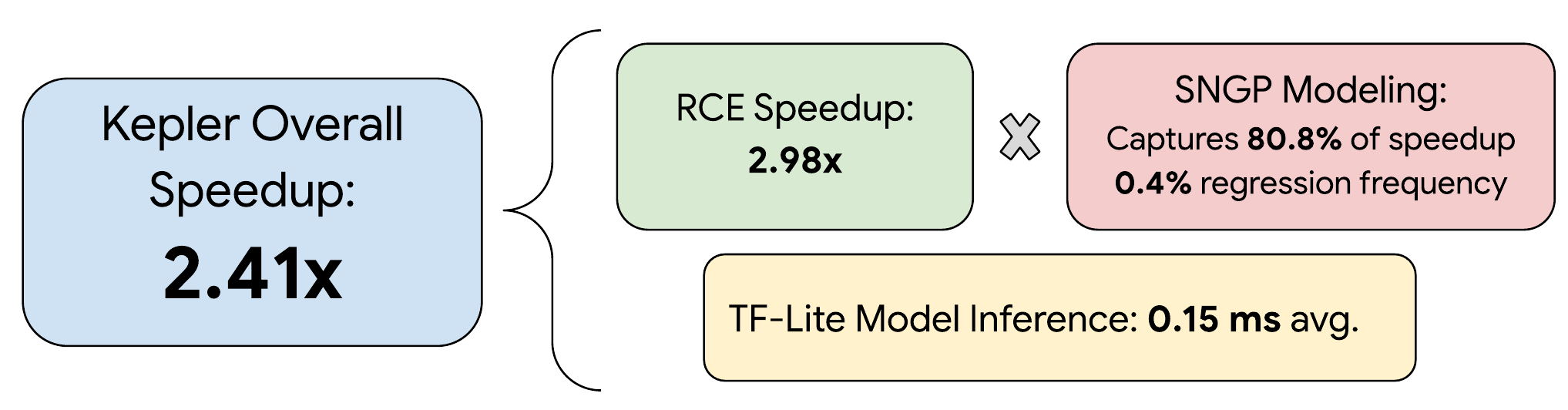}
    \caption{Kepler achieves an overall 2.41x speedup on Stack by 1. discovering better plans via RCE, 2. capturing the majority of the speedup with SNGP models while minimizing regressions, and 3. having fast model inference time.}
    \label{fig:overall_results}
\end{figure}

We propose that restricting the query optimization problem to the parameterized query setting poses a more tractable learning problem and hence can be more robustly solved. To this end, we present \sysname (\textbf{K}-plan \textbf{E}volution for \textbf{P}arametric Query Optimization: \textbf{L}earned, \textbf{E}mpirical, \textbf{R}obust), an end-to-end learning-based approach for parameterized queries. Building on prior work in PQO \cite{vaidya2021leveraging}, \sysname leverages a novel plan generation strategy, a training query execution phase, and a robust neural network model design. Combined, we show that these techniques provide significant improvements in both planning time and query execution performance, satisfying both the PQO and QO objectives. Best of all, \sysname's use of robust neural network techniques drastically reduces the frequency and magnitude of performance regressions. Figure~\ref{fig:overall_results} highlights how each of \sysname's components contribute to a 2.41x geometric mean speedup across the entire Stack benchmark \cite{marcus2021bao}.

\sysname follows a decoupled plan generation and learning-based plan prediction architecture similar to the approach of \cite{vaidya2021leveraging} with three key differences. First, \sysname provides the key insight that designing better candidate plan generation algorithms can lead to substantially faster plans than the built-in optimizer's. We propose Row Count Evolution (RCE), a method that efficiently generates candidate plans by perturbing the optimizer's cardinality estimates. RCE only requires a simple interface to any standard cost-based optimizer, making it compatible with most database systems. 

Second, \sysname leverages actual query execution data to build a training dataset for best-plan prediction, avoiding the well-studied mismatch between cost models and execution latency~\cite{leis2015good}. While \sysname's collection of execution data may be costly if the parameterized query is run infrequently, we argue that the additional execution data in our setting is justified by (1) the scale of parameterized queries in production and (2) the query execution speedups afforded by RCE.

Third, \sysname uses robust neural network prediction techniques to decrease tail latency and reduce query regressions (i.e. worse performance than the existing query optimizer). Specifically, \sysname uses Spectral-normalized Neural Gaussian Processes (SNGPs)~\cite{liu2020simple} to accurately quantify how confident it is about a prediction, and falls back to the database's query optimizer when it is uncertain.

\paragraph{Our contributions}
\begin{itemize}
    \item{We identify a novel and practical formulation of query optimization for parameterized query templates in which \emph{speedups} against a classical query optimizer can be robustly achieved.}
    \item{We propose a novel candidate plan generation algorithm, Row Count Evolution (RCE), that produces significant speedup compared to classical query optimizers on real-world and synthetic datasets.}
    \item{We demonstrate that incorporating robust ML techniques allows models to capture large portions of the speedups while greatly reducing the risk of regressions.}
    \item{We demonstrate that our model inference costs are negligible via an end-to-end PostgreSQL integration for the query path.}
    \item{We open-source both our system implementation for PostgreSQL\footnote{https://github.com/google/kepler} as well as our query execution datasets, which we believe is the first dataset tailored towards parameterized query optimization. The datasets collectively represent $\sim$14.2 CPU years of query execution time. They serve as a benchmark for further work on best-plan prediction as well as simulating more efficient techniques for training data collection.}
\end{itemize}

%% file: sections/2_related_work.tex
\section{Related Work}
\label{sec:related}

\textbf{Parametric query optimization.}
PQO has been extensively studied in a variety of works~\cite{alucc2012parametric, hulgeri2002parametric, ioannidis1997parametric, vaidya2021leveraging, chaudhuri2010variance, dutt2017leveraging}. The goal of the standard PQO formulation is to reduce the amount of times the query optimizer is invoked while minimizing the corresponding regression in query latency~\cite{dutt2017leveraging, vaidya2021leveraging, alucc2012parametric}. Although \sysname also focuses on parametric queries, its primary objective is closer to that of standard query optimization, which seeks to improve query latencies. \sysname also simultaneously improves on the PQO objective by leveraging fast-inference ML models.

Prior PQO approaches typically make simplifying assumptions such as heavily relying on the optimizer cost model or using base table selectivities as input features~\cite{vaidya2021leveraging, dutt2017leveraging}. This may be feasible for some advanced commercial systems; however, this over-reliance on the existing optimizer is particularly dangerous given the well-studied deficiencies of optimizers such as PostgreSQL~\cite{leis2015good}.

Our approach follows a similar structure as~\cite{vaidya2021leveraging}, which also decouples the populateCache (candidate generation) and getPlan stages (ML-based prediction). However, since they focus on the standard PQO objective of attempting to match the existing optimizer, they require using a bandit algorithm to reduce their training data cost. By contrast, the primary objectives of \sysname are query performance and robustness, leading to a lower emphasis on training query efficiency.

Several popular database systems have implemented PQO features, including Oracle Adaptive Cursor Sharing, Aurora Managed Plans, and SQL Server Parameter Sensitivity Plan optimization~\cite{aurora,oraclecursor,sqlserverpsp}. These features all heavily rely on their cost models (based on traditional statistics and heuristics), and do not utilize machine learning models. 
\paragraph{Query plan generation.} 
\rewrite{Several prior works suggest methods for candidate generation, which we divide into four main categories.}
\begin{enumerate}
\item\textbf{Default optimizer plans.} The simplest method combines the optimizer's selected plan for each query instance. This approach is frequently found in PQO algorithms since they seek to cache the optimizer's plans~\cite{vaidya2021leveraging}. This strategy is also employed in~\cite{snodgrass2022have} to estimate the empirical suboptimality of existing query optimizers.
The quality of the resulting candidate plan set is predicated upon the optimizer's ability to either generate optimal plans for each query instance or a sufficient variety of good plans across the workload to benefit from plan sharing. However, we empirically observed that the optimizer fails to do so on real-world datasets. (Table~\ref{tab:pruning}).

\item\textbf{Cost-based plan pruning.} populateCache algorithm ~\cite{vaidya2021leveraging} extends the default optimizer candidate generation method with cost-based $K$-set identification to prune the candidate set to size $K$. However, this pruning method may mistakenly prune good plans if the correlation between the cost estimates and actual execution times are poor. 

\item\textbf{Optimizer configuration parameters.} Query optimizers typically expose a variety of configuration parameters that can be used to alter their query planning behavior. In particular, PostgreSQL has configuration parameters that allow one to disable entire classes of join and scan operators from being used in query plans. Bao selectively applies subsets of these parameters in order to generate new query plans~\cite{marcus2021bao}. Although simple, disabling operator types is a heavy-handed and indirect approach to generating new plans.

\item\textbf{Exact cardinalities.} Exact cardinality query optimization (ECQO) attempts to construct the optimal plan by computing the plan induced by the exact cardinality values of all possible sub-plans~\cite{chaudhuri2009exact}. However, for sufficiently complex queries, evaluating these exponentially-many sub-plans is prohibitively slow even with optimizations~\cite{trummer2019exact}. The selected plans are also not always the fastest, as observed by~\cite{snodgrass2022have}.
\end{enumerate}

In summary, these methods are all unsatisfactory for a variety of reasons: failure to generate faster plans (1, 2, 4), ineffectively exploring the plan space (3), or are computationally intractable (4).

\paragraph{Machine learning for query optimization.}
A wide range of techniques apply ML on QO, most notably for predicting cardinality estimates (CE)~\cite{yang2020neurocard, yang2019deep, kipf2018learned}. Recent work show cardinality estimation may be brittle in practice, and that even small Q-errors can lead to noticeably worse plans~\cite{leis2015good, wang2020we}. In general, these work do not measure the actual end-to-end execution latency of selected plans after integrating their models into an optimizer~\cite{lu2021pre}.

Several approaches have demonstrated improved query performance, but typically do not consider the issue of robustness. Neo~\cite{marcus2019neo} and Bao~\cite{marcus2021bao} leverage tree convolutional neural networks to adaptively optimize plans using reinforcement learning and contextual bandits respectively. These online algorithms offer no guarantees on stability or regression avoidance, and hence cannot reliably be deployed in production. Similarly, techniques applying deep reinforcement learning to QO have not demonstrated consistently better performance and suffer from robustness issues~\cite{krishnan2018learning, marcus2018towards, yang2022balsa}. For example, Figure 9 in~\cite{yang2022balsa} indicates a significant amount of regressions both at train and test time.

%% file: sections/3_overview.tex
\section{Overview}
\label{sec:overview}
In this section, we describe our problem setting (Section~\ref{ss:setting}), give an overview of our approach (Section~\ref{ss:system-overview}), and further discuss specific design choices that are made in \sysname (Section~\ref{ss:design-choices}).

\subsection{Problem Setting}
\label{ss:setting}
As in prior work~\cite{vaidya2021leveraging}, we consider parameterized queries that are repeatedly invoked with different parameter bindings. Such queries are specified by a template $Q$ with $m$ parameterized predicates\footnote{The parameters do not necessarily have to be in predicates, e.g. they may appear in a LIMIT clause. However, our experiments only include the parameterized predicate case.} $x_0,\ldots ,x_{m-1}$ of varying data types. We let $q$ denote a specific query instance, i.e. $Q$ with a fixed set of parameter binding values. A query plan $p$ associated with a template $Q$ specifies how to execute any query instance $q \sim Q$. A \emph{sub-plan query} of $Q$ is $Q$ restricted to only a subset of its tables \cite{han2021cardinality}, and its output cardinality is referred to as its \emph{sub-plan cardinality}. We assume a fixed database system with a built-in query optimizer, and denote the default plan $p_{\texttt{default}}(q)$ to be the plan selected by the query optimizer for $q$. Finally, a workload $W \subset \mathcal{W}$ consists of a set of query instances $\{q_0, \ldots, q_{n-1}\}$ for a single template $Q$, where $\mathcal{W}$ denotes the space of all possible query instances.

\subsection{\sysname Overview}
\label{ss:system-overview}

Our approach at a high level follows that of \cite{vaidya2021leveraging}: we consider a single, isolated query template $Q$, and decouple the problems of generating a set of possible plans and deciding which plan to use for each query instance. More formally, these problems can be described as:
\begin{enumerate}
    \item \textbf{Candidate generation.} Generate a candidate set of $k$ plans $\{p_0, \ldots, p_{k-1}\}$ for $Q$, out of the exponentially-large set of all possible plans $\mathcal{P}$ (corresponding to \texttt{populateCache} in~\cite{vaidya2021leveraging}).
    \item \textbf{Best-plan prediction.} Learn a mapping $M: \mathcal{W} \rightarrow P$ that minimizes some objective, e.g. some measure of execution latency over the workload (corresponding to \texttt{getPlan} in~\cite{vaidya2021leveraging}).
\end{enumerate}

Unlike~\cite{vaidya2021leveraging}, who attempt to match the performance of the built-in optimizer, our goal is to improve upon the built-in optimizer as much as possible. To achieve this, \sysname includes a sophisticated candidate generation algorithm, described in Section~\ref{sec:candidate_generation}, that empirically generates better plans than the built-in optimizer. The afforded speedups allow \sysname to avoid relying on potentially-brittle online learning approaches (e.g. contextual bandits) during the training data collection phase.

\paragraph{Objective.} We first define several key metrics and terms in our problem setting. For a given query instance, we denote the optimal plan over some plan set $P$ as $p^{P}_{\texttt{opt}} = \min_{p\in P}{ExecTime(p, q)}$, where $ExecTime$ refers to the actual execution time. We define $p^*_{\texttt{opt}}(q)$ as the optimal plan over all possible plans, i.e. when $P=\mathcal{P}$. Typically, the optimal plan refers to $p^*_{\texttt{opt}}$ for candidate generation and $p^{P}_{\texttt{opt}}$ for modeling. We also refer to near-optimal plans as plans that have similar execution time to $p^{P}_{\texttt{opt}}$ or $p^*_{\texttt{opt}}$.

For some fixed candidate set $P$, we define the (oracle) speedup ratio relative to the default plan as:
\begin{equation}
\label{eq:oracle_speedup}
    S^{\texttt{opt}}(P, W) = \frac{\sum_{q\in W} ExecTime(p_{\texttt{default}}, q)}{\sum_{q\in W} ExecTime(p^{P}_{\texttt{opt}}, q)}
\end{equation}
 This quantity is the factor by which we can improve the total execution time of the workload if we had oracle access to the optimal plan in $P$ for each query instance. We note that this ratio corresponds exactly with the definition of execution cost sub-optimality in \cite{vaidya2021leveraging}; the re-naming to speedup emphasizes the differences in our system objectives. Since we can union $P$ with the set of all default plans over $W$, this speedup ratio is always lower bounded by 1. 

Similarly, for some model $M: \mathcal{W} \rightarrow P$, we define its model speedup as:
\begin{equation}
\label{eq:model_speedup}
    S^{\texttt{model}}(W) = \frac{\sum_{q\in W} ExecTime(p_{\texttt{default}}, q)}{\sum_{q\in W} ExecTime(p_{\texttt{model}}, q)}
\end{equation}
This quantity corresponds to how much faster the model is at executing a workload than the default optimizer. Although $S^{\texttt{model}}$ is by definition upper bounded by $S^{\texttt{opt}}$, it is not necessarily lower bounded by 1, i.e. if the model selects plans worse than the default plan.

An auxiliary objective of Kepler is reducing workload tail latency. Several work have identified that database optimizers may perform significantly worse in the tail of the query latency distribution, which poses a significant obstacle for use cases that require a more uniform runtime~\cite{marcus2021bao}.

\paragraph{\sysname architecture.}

\begin{figure}[!t]
\includegraphics[width=\linewidth]{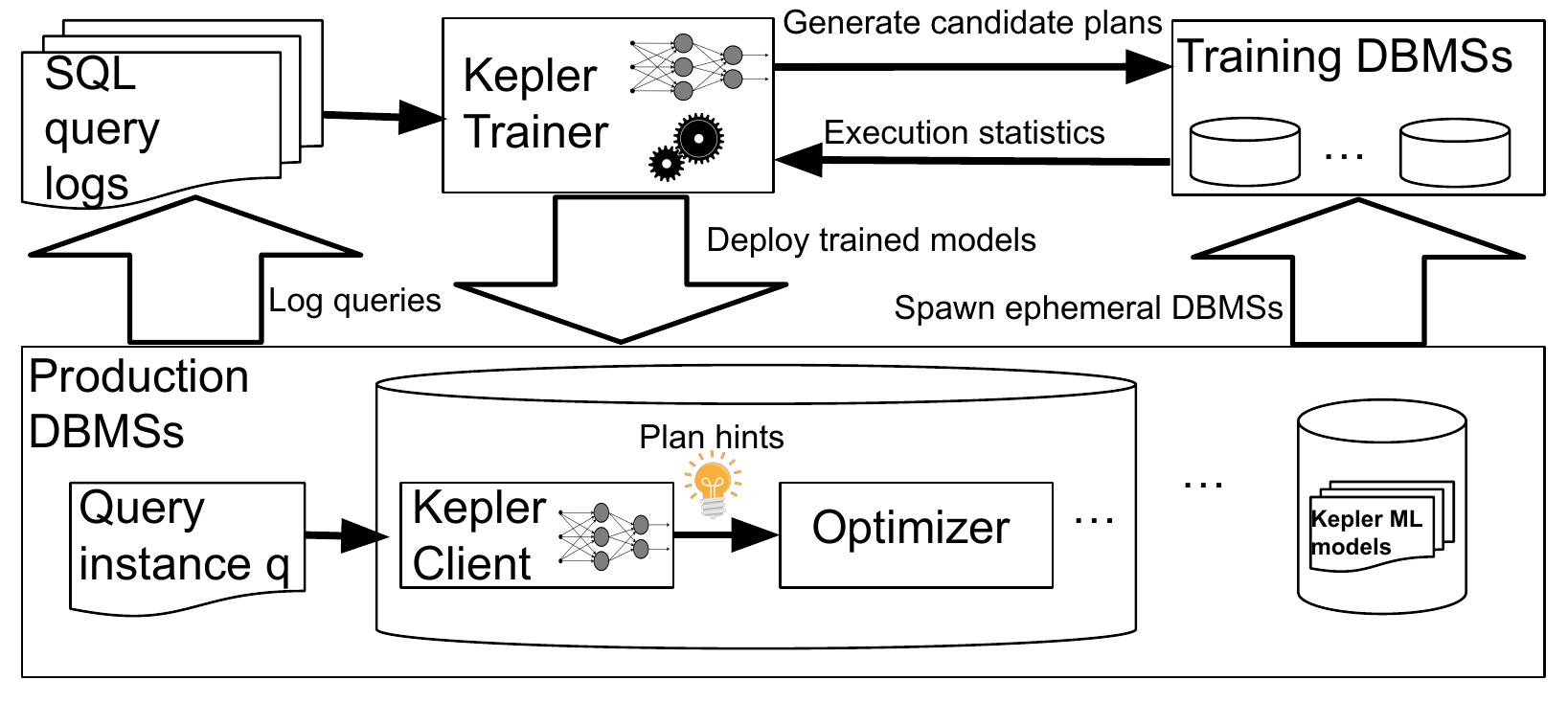}
\caption{\sysname architecture.}
\label{fig:design}
\end{figure}

Figure~\ref{fig:design} shows the architecture of \sysname, consisting of a \sysname trainer and \sysname client. The trainer ingests query instances from the query logs produced by production DBMSs and aggregates them into query templates. For each query template $Q_i$, the \sysname trainer aims to find the near-optimal plans for all its query instances ${q_j}$. It uses Row Count Evolution (RCE) to generate candidate plans ${p_k}$ and executes the queries with these plans to collect execution statistics. To minimize impact on production DBMSs, the trainer may optionally request a production DBMS to spawn ephemeral instances to execute these queries. The \sysname trainer trains an ML model to predict the best plan for $q_j$ based on these execution statistics and deploys the trained models into the production DBMSs. 

A \sysname client maintains a mapping from a query template to an ML model. When a production DBMS receives a query instance $q$, the client first checks if an ML model is available for $q$. If available, it performs model inference to predict the best plan hints and provides the hints to the optimizer only if the associated confidence score is higher than a threshold. Otherwise, it falls back to the built-in optimizer to produce a plan. 

\paragraph{Changing environments and workloads.} In our current implementation, \sysname assumes a fixed system state, including database configuration, optimizer implementation, and data distribution. If any of these aspects changes relatively slowly or infrequently, \sysname can periodically collect new execution data and retrain purely on data from the new system state. \rewrite{We posit that in the majority of production parameterized query use cases, (1) the database is reconfigured infrequently, and (2) the data distribution drifts slowly, e.g. in scenarios in which a relatively small amount of similarly-distributed data is added each day.} Additionally, \sysname is designed to be robust to dynamic workloads in which query parameter binding values change by detecting when inputs are out of its training distribution (see Section~\ref{subsec:model_exp}).

\paragraph{Limitations.} The target usage of \sysname is for parameterized queries that are executed frequently enough to justify the training data collection cost. \rewrite{As discussed in Section~\ref{sec:future_work}, the exact training data collection regime in this paper serves the dual purposes of definitively demonstrating the speedups available and enabling further research in efficiency. We anticipate a final production system will use an iteration of this research with leaner training data collection.} The cost-benefit analysis of using \sysname is situation-dependent; ultimately the user must weigh the potential query performance gains against the cost. If ephemeral instances are used for training data collection, \sysname assumes they are representative of production query performance.

\subsection{\sysname Design Choices}
\label{ss:design-choices}
In this section, we further discuss the specific design choices made to ensure that \sysname can be reliably deployed with minimal production overhead.

\paragraph{Using actual execution latencies.} Since the objective of \sysname is to reduce actual end-to-end query latencies, it necessitates executing queries on a real database to provide ground-truth signal. To minimize the training collection time and avoid load on the production system, the DBMS may spawn ephemeral instances to speed up and isolate the training execution process.

\begin{figure}[!t]
    \centering
    \includegraphics[scale=0.4]{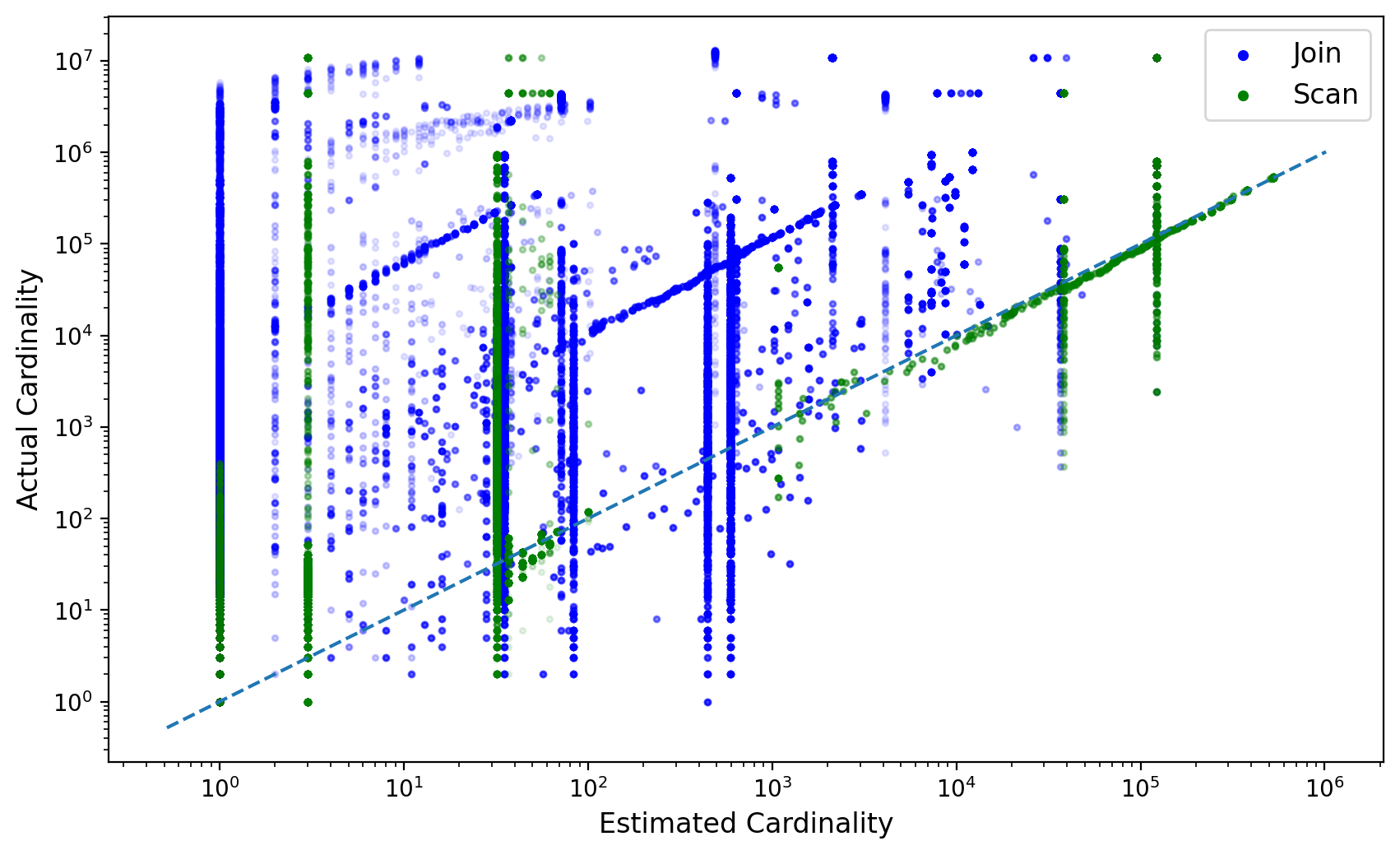}
    \caption{Predicted vs. exact cardinalities on instances of Stack q12\_2. Dashed y=x line is ideal (i.e. actual = estimated).}
    \label{fig:pred_vs_actual_q12_2}
    \vspace{-1em}
\end{figure}

\paragraph{Limiting reliance on cost models.} By collecting actual execution latencies, \sysname eschews explicitly relying on optimizer cost estimates for determining the quality of a plan. Figure~\ref{fig:pred_vs_actual_q12_2} shows the estimated vs. exact cardinalities of all joins on a sample of query instances from Stack \cite{marcus2021bao}. In particular, 64\% of points have estimated cardinality = 1, likely due to the independence assumption of the PostgreSQL optimizer.
 
\paragraph{Falling back to the built-in optimizer.} \sysname avoids regressions by falling back to the existing query optimizer when it is not confident in identifying the optimal plan. Given the low overhead of model inference, the overall \sysname inference cost is nearly always lower than that of Opt-Always. For cases where planning time is a concern due to high fallback frequency, one can incorporate an additional model designed to predict a safe plan without re-invoking the optimizer.

\paragraph{Independence of query templates.} \sysname handles query templates independently, i.e. each query template will generate its own candidate plans, collect its own training data, and train a model specific to that template. Though potentially more expensive than a procedure that generalizes over multiple query templates, this design has the advantages of 1) providing a more tractable ML problem, and 2) isolating each query from regressions caused by changes pertaining to other queries as models iterate over time and new query templates are on-boarded. Leveraging shared information between query templates while not increasing the risk of regressions is an interesting direction for future work.

%% file: sections/4_candidate_generation.tex
\section{Row Count Evolution} 
\label{sec:candidate_generation}
\label{subsec:row count evolution}
The goal of the candidate generation stage is to construct a set of plans $P$ such that it contains a near-optimal plan for every query instance $q$ in the workload distribution $\mathcal{W}$. Additionally, $P$ should be sufficiently small such that it is feasible to execute each plan for training dataset collection. Balancing these two competing objectives is the main challenge for any candidate generation algorithm.

In this work, we only consider generating fully-specified plans, i.e. the join order and every join/scan method are defined. Alternatively, a candidate generation algorithm could specify a subset of the plan decisions and allow the query optimizer to determine the remainder.

\paragraph{Workload candidate generation.} Given an algorithm $A$ for generating a candidate plan set over a single query instance $q$, we define the corresponding plan set over a workload $W$ as the union of the per-instance plan sets $A(W) := \bigcup_{q\in W} A(q)$ (see lines 1-5, Algorithm~\ref{alg:rce}). We also define \emph{plan sharing} to describe the case where $p^P_{\texttt{opt}}(q)$ is generated from some other query instance $q'$ (i.e. $p^P_{\texttt{opt}}(q) \notin A(q)$, $p^P_{\texttt{opt}}(q) \in A(q')$).



\paragraph{Our approach.} We propose Row Count Evolution (RCE)\footnote{The name "Row Count" is inspired by the PostgreSQL extension pg\_hint\_plan's row count hints, which we use to modify the PostgreSQL optimizer's cardinality estimates.}, a computationally-efficient algorithm that generates new plans by randomly perturbing the optimizer's cardinality estimates. RCE is predicated on the idea that cardinality misestimates are the primary underlying reason for optimizer suboptimality. RCE exploits the fact that our candidate generation stage only needs to generate a set of plans that contains a (near-)optimal plan instead of directly identifying a single performant plan. Like Bao~\cite{marcus2021bao}, RCE leverages the built-in query optimizer to generate candidate plans, but does so in a more fine-grained and efficient way.


\begin{table}[t]
\small
\centering
\caption{\rewrite{RCE hyperparameters.}}
\begin{tabular}{ll} \hline
Notation & Definition \\ \hline
$G$ & Number of generations. \\ \hline
$b$ & Exponent base for row count perturbation. \\ \hline
$m$ & Exponent range for row count perturbation. \\ \hline
$S$ & Number of plans sampled from the previous generation. \\ \hline
$N$ & Number of perturbation for each candidate plan. \\ \hline
\end{tabular}
\label{tab:term}
\end{table}

\begin{algorithm}[t]
\caption{\rewrite{Row Count Evolution.}}\label{alg:rce}
\begin{algorithmic}[1]
\label{alg:workload_gen}
\Function{WorkloadCandidateGeneration}{workload $W$}
    \State $P \gets \{\}$
    \For {query instance $q \in W$}
        \State $P \gets P \cup \text{RowCountEvolution}(q)$
    \EndFor
    \State \Return $P$
\EndFunction

\State
\Function{RowCountEvolution}{query instance $q$}
    \State $p_0=$ the base plan for $q$
    \State $C_0 = \{(p_0, \{\}, \{s\rightarrow 0\, \forall \text{ sub-plans } s\})\}$
    \For {generations $g = 1, 2, \ldots G$} 
        \State Sample up to $S$ base plans $B_g$ uniformly from $C_{g-1}$
        \State $C_g \gets \{\}$
        \For {(base plan $p$, row count map $r$) $\in B_g$}
            \For {$i = 1, 2, \ldots, N$}
                \State $r' \gets$ SamplePerturbations($p$, $r$) 
                \State $p' \gets$ GetOptimizerPlan($r'$)
                \If {$p' != p$}
                    \State $C_g$.add(($p'$, $r'$))
                \EndIf
            \EndFor
        \EndFor
    \EndFor
    \State \Return $C_0 \cup C_1 \cup \ldots \cup C_G$
\EndFunction

\State
\Function{SamplePerturbations}{plan $p$, row count map $r$}
    \For {sub-plan $s \in p$}
        \State $w \gets p$.getEstimatedCardinality($s$)
        \State $e_l \gets -\min(\log_{b}(w), m)$
        \State $e_u \gets e_l + 2m$
        \State Sample $f$ uniformly from $[b^{e_l}, \ldots, b^{e_u}]$
        \State $r[s] \gets w\cdot f$
    \EndFor
    \State \Return $r$
\EndFunction
\end{algorithmic}
\end{algorithm}

We instantiate the idea of applying random perturbations as an evolutionary-style algorithm, described in Algorithm~\ref{alg:rce}. RCE maintains a sequence of generations of plans, with the initial generation consisting solely of the query optimizer's plan. To construct subsequent generations, RCE first uniformly samples parent plans from the previous generation. For each of these base plans, RCE perturbs the \rewrite{join} cardinalities of \emph{only the sub-plans that appear in the parent plan} by multiplicative factors sampled from an exponentially-spaced range (lines 21-28). By repeating this process multiple times and feeding in the resulting perturbations into the query optimizer, RCE generates a set of children plans (lines 14-18). Out of these, only unseen plans (i.e. those that did not appear in any prior generation) are kept for the next generation (lines 17-18). 

\begin{figure}[!t]
\includegraphics[width=\linewidth]{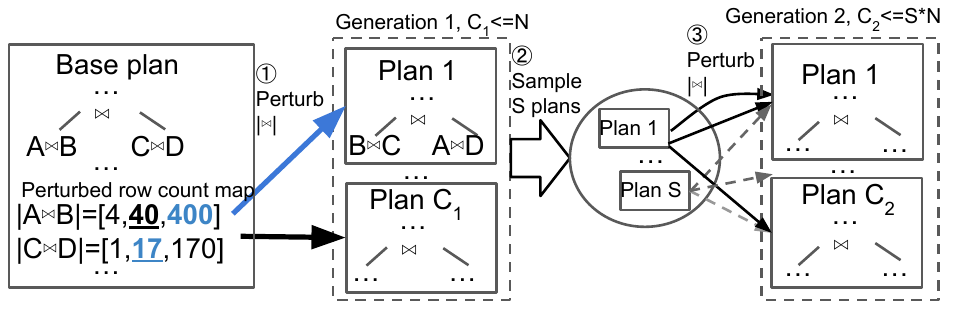}
\caption{\rewrite{An example RCE process.}}
\label{fig:rce_example}
\vspace{-1em}
\end{figure}

\rewrite{
\paragraph{Example.}
Figure~\ref{fig:rce_example} shows an example of RCE generating candidate plans for a query instance with two generations. 
The base plan joins the result of $A\Join B$ and $C\Join D$ with estimated $|A\Join B|=40$, $|C\Join D|=17$. 
RCE first constructs a set of candidate row counts for each sub-plan by perturbing their cardinalities by multiplicative factors. These candidate row counts for $A\Join B$ and $C\Join D$ are $[4, 40, 400]$ and $[1, 17, 170]$, respectively, using a base of 10 and a range of 1.
RCE then uniformly samples new join cardinalities from these sets; one sample of 400 and 17 influences the optimizer to produce a new plan Plan-1 in generation 1~\circled{1}. 
It repeats the same process N times to produce $C_1$ plans in generation 1. 
Next, RCE samples $S$ from a deduplicated set of plans from generation 1~\circled{2} and randomly perturbs the row counts on each sampled plan $N$ times to generate $C_2$ plans in generation 2~\circled{3}. 
}

\paragraph{RCE as exact cardinality matching.} One interpretation of RCE is that it efficiently builds a covering set of exact-cardinality plans. The RCE-generated candidate set contains plans generated from a diverse range of perturbed sub-plan cardinalities. If there are sufficiently many perturbations, likely at least one will be reasonably close to the exact cardinalities for any particular query instance and their respective induced plans will also likely be similar.

\paragraph{Multiplicative perturbations.} Applying multiplicative perturbations is well-motivated by the standard metric of Q-error in cardinality estimation. RCE further uses an exponentially-spaced perturbation set in order to have a similar support as the optimizer's Q-error distribution.

\paragraph{Perturbing only relevant sub-plans.} Instead of perturbing all $2^n - 1$ sub-plans (for a query joining $n$ tables), RCE only perturbs the cardinalities of the $n-1$ sub-plans that actually appear in the sampled query plans. This significantly increases the efficiency of RCE with only a small loss of generality: since the set of perturbations is inherited between generations, a misestimated sub-plan cardinality will only never be perturbed if its cardinality is significantly overestimated by the query optimizer. However, this is an unlikely scenario since query optimizers tend to underestimate sub-plan cardinalities due to the independence assumption. 


This re-optimization of only the sub-plan cardinalities that appear in the optimizer plan bears a strong resemblance to the re-optimization procedure in \cite{wu2016sampling}, which iteratively re-optimizes using sampling-based cardinality estimates. The key differences in our setting are (1) we do not have to return a single plan, and (2) we require a fast procedure since we repeat it for each query instance, motivating the use of perturbations over sampling.

\paragraph{RCE as local search.} RCE effectively explores the plan space via a random walk in the low-dimensional subspace of sub-plan cardinalities, initialized at the optimizer's cardinality estimates. This formulation implicitly leverages the fact that while these initial estimates are typically incorrect, they are still more informative than random estimates.

\paragraph{RCE hyperparameters.} Our implementation of RCE includes a variety of hyperparameters that allow one to flexibly trade off the number of generated plans against the potential total speedup (Table~\ref{tab:term}). 

\begin{itemize}
    \item \textbf{Width and depth of the perturbation tree.} \rewrite{Increasing the number of generations $G$ increases the number of plans, making it more likely a good plan is found. However, plans in later generations are perturbed further from the original plan, and may have a lower likelihood of being relevant. To ensure constant-time processing for each generation, we sample (up to) a fixed number $S$ of base plans in each generation, and perturb each one $N$ times.}
    \item \textbf{Perturbation values.} \rewrite{The exponent base $b$ and range $m$ limits the magnitude of a single perturbation.} We also introduce a sub-plan perturbation limit that controls the number of times a specific sub-plan can be perturbed, effectively controlling the total perturbation range of any given sub-plan.
    \item \textbf{Direct limits on number of plans.} We implement limits on the number of plans that can be generated from a single parameter and in total. Once the limit is reached, the evolutionary candidate generation process is terminated and only default plans are kept for remaining parameters. \rewrite{The total plans limit is a soft limit since the final evolutionary iteration may produce up to the single-parameter limit and the remaining parameters may contribute new default plans.}
\end{itemize}

%% file: sections/5_training_execution.tex
\section{Training Data Collection}
\label{sec:training_data}

After generating candidate plan set $P$, we execute each plan over a training workload to generate a dataset of execution latencies for supervised best-plan prediction. The training workload may be provided by the user or captured in a DBMS query log ~\cite{vaidya2021leveraging}. Rather than executing all candidate plans for each query instance, we use adaptive timeouts and construct near-optimal plan covers to prune suboptimal plans.


\paragraph{Execution mechanics.} We force the optimizer to produce a candidate plan by providing all join/scan methods and the join order via hints. 
We parallelize the execution of query instances and their candidate plans in multiple databases. 
We simulate a warm buffer cache scenario by executing each plan multiple times and taking the minimum as the estimated latency~\cite{leis2015good}. 
This repeated execution strategy also reduces the potential noise in our execution time measurements; though we observed the amount of noise to be inconsequential in our experimental setup. 
We leave a full analysis of query execution time under different caching, concurrency, and resource availability settings to future work.

\paragraph{Adaptive timeouts and plan execution reordering.}
We use a timeout policy to minimize wasted resources on executing sub-optimal candidate plans. 
The timeout policy adapts from ~\cite{yang2022balsa} with two main modifications. 
First, we always execute the default plan first and adaptively reorder the remaining plans to maximize the impact of the timeout's progressive tightening on a per query-instance basis. 
We execute plans in ascending order of their historical execution latencies across query instances as a simple heuristic for tightening the timeout as quickly as possible. 
Second, we do not apply the tightened timeout for the first iteration of each plan in order to ensure that each plan simulates a warm-cache scenario. 

\paragraph{Online plan cover pruning.} 
We also use an online plan pruning technique to eliminate plans based on actual execution time. 
Specifically, we initially execute all plans for the first $N$ query instances of a query template, then use a Set Cover formulation to prune down to a minimal \emph{plan cover} set for the remaining query instances. 
The pruned set becomes our $k$ candidate plans for the query template, i.e. our models only attempt to predict from those plans.

We consider a plan to be near-optimal for a query instance $q$ if its execution time is within a $1 + \epsilon$ factor of the fastest time for $q$ we have seen so far. 
Each plan has an associated set of query instances for which it is near-optimal. 
The \emph{plan cover} is the smallest set of plans such that each query instance has a near-optimal plan in the set. 
We construct the plan cover using the standard greedy approximation for Set Cover, which iteratively picks the plan that is near-optimal for the most remaining query instances. 
We additionally relax the problem to require that only $1-\delta$ of all query instances be covered, allowing us to trade off the plan cover size and the achievable speedup. 

\paragraph{Tail latency reordering.} For many query templates, the distribution of default execution latencies is heavy-tailed. Parameters in the tail tend to be more sub-optimal, and therefore have an outsized impact on the total speedup. 
To ensure the plan cover computation includes these parameters, we evaluate these query instances first.
This reordering produces a 7-8x reduction in total execution time and number of plans over the entire Stack dataset.


%% file: sections/6_modeling.tex
\section{Robust Best-Plan Prediction}
\label{sec:modeling}

After collecting a full training dataset of actual execution latencies over our candidate plan set, we use supervised ML to predict the best plan for any query instance. 
\sysname trains one model for each query template with the objective to maximize workload speedup while minimizing regressions. 
\sysname also falls back to the optimizer's plan when its predicted confidence is low. Section~\ref{sec:experiments} shows that the inference time of our model is negligible compared to the typical query planning time of a classical optimizer.

\subsection{Features}

Given a template $Q$ with $m$ parameters, \sysname uses solely the $m$ parameter values as input features. 
The supported types include numerics (float/int), strings, and dates/timestamps. 
We apply standard preprocessing techniques to each type: embeddings for strings/low-dimensional integer features, normalization to $N(0, 1)$ for numerics, and numeric conversion for date/time features.

We do not  convert the parameter values to their respective base table selectivities as in \cite{vaidya2021leveraging} for the following reasons. First, selectivity is inherently a lossy representation and may obscure information when two distinct values have the same selectivity. Second, selectivity is inferior when the optimizer's cardinality estimation is sub-optimal, see Figure~\ref{fig:pred_vs_actual_q12_2}.

\paragraph{String columns and vocabulary selection.} For each string-valued column, we construct a fixed-size vocabulary in order to limit model size. String features are one-hot encoded via a lookup table, with buckets for out-of-vocabulary values. An embedding layer is then applied on this one-hot encoding, creating a learnable embedding for each value in the vocabulary.

We choose the vocabulary as the top-$k$ values ordered by the total possible improvement of all query instances with that value. We define the \emph{max marginal improvement} strategy as selecting the top-$k$ column values $v$ in column $i$ under the following objective:
\begin{equation}
    m(v, i) = \sum_{q\in W, x_i = v} ExecTime(p_{\texttt{default}}, q) - ExecTime(p^P_{\texttt{opt}} , q)
\end{equation}

Our evaluation shows that this strategy is effective. For columns with significantly more distinct values, one may factorize embeddings over subcolumns~\cite{yang2020neurocard}.

\subsection{Training Objectives}
\label{subsec:training_objectives}
\sysname models maximize the model speedup defined in Equation~\ref{eq:model_speedup} while minimizing the number of regressions. 
This objective is not directly differentiable, so we discuss various surrogate learning objectives. 

\paragraph{Multi-label classification.}
We model best-plan prediction as a \emph{multi-label} classification problem in which each near-optimal plan has a positive label (as opposed to just the optimal plan) \cite{tsoumakas2007multi}. The multi-label objective also provides a richer supervised signal, improving the quality of the learned intermediate representations. We use the single-label transformation of multi-label classification loss by training the near-optimal probability of each candidate plan with binary cross-entropy loss.

Although our models only predict plans in the plan cover, which may not necessarily contain every query instances' default plan, our definition of near-optimality does exploit the availability of default plan execution data during training. We define a plan to be \emph{near-optimal} if its estimated latency improvement is within a $1 + \tau$ factor of the optimal improvement latency. Namely, we say a plan $p$ is near-optimal if $(\ell_d - \ell_p)(1+\tau) \geq (\ell_d - \ell_o)$, where $\tau > 0$, $\ell_d = ExecTime(p_{\texttt{default}}, q)$, $\ell_p = ExecTime(p, q)$, and $\ell_o = \min_{p\in P}{ExecTime(p, q)}$. \rewrite{Computing near-optimality requires execution times for all query instances for all plans.}

Prior work formulate best-plan prediction as a regression~\cite{vaidya2021leveraging, marcus2018deep, akdere2012learning} and multi-class classification problem~\cite{vaidya2021leveraging}. Both formulations are unsatisfactory for a variety of reasons. Regression across a significant range can be unstable, a problem that is exacerbated by our timeout procedure, which obscures the true latency of suboptimal plans. \rewrite{Regression attempts a more challenging problem with finer granularity than required, imposing unnecessary constraints and objectives on the training. We only need to predict the identity of the optimal plan rather than its execution time. Inversions and gross over-estimates of non-optimal plans are acceptable to us but will weigh heavily in regression loss.} 
Meanwhile, classification objectives that predict a single optimal plan perform poorly in scenarios when multiple plans can be near-optimal and empirical execution latencies can be subject to noise. For example, consider a problem where plans $p_1, p_2$ execution latencies' are both drawn from $C + N(0, 1)$ for some large $C$. Then a multi-class classifier will have equal predicted likelihood for $p_1$ and $p_2$ and thus have low confidence, when in actuality being confident in $p_1$ and/or $p_2$ is desirable.

\paragraph{Example-dependent loss.} Different query instances may have disproportionate impact on the overall objective Equation~\ref{eq:model_speedup}. We leverage the standard sample-weighting approach example-dependent cross entropy \cite{hepburn2018proper, bahnsen2014example} to prioritize those with the largest improvement delta. For plans worse than the default plan, we upweight them by a factor $C$. For all near-optimal plans, we apply a soft weighting based on their empirical execution improvement, i.e. $1 + D\log(\ell_d - \ell_p)$, where $C$ and $D$ are both tunable hyperparameters. 

\subsection{Models}
We use simple feedforward neural networks as our base models. For inference efficiency, we consider a neural network with one output head per plan on top of a shared representation, which improves inference speed and model size over approaches that have separate models for each plan~\cite{vaidya2021leveraging}. 

We train our neural network models with standard minibatch SGD. In a real-world setting, the model's hyperparameters can be tuned via simple search techniques or more sophisticated algorithms by partitioning the training data into a train and validation set.

\paragraph{Uncertainty.} \sysname models incorporate calibrated predictions and uncertainty estimates to avoid predicting significantly suboptimal plans. Two state-of-the-art approaches for incorporating uncertainty into neural networks are ensembling and Spectral-normalized Neural Gaussian Processes (SNGPs) \cite{liu2020simple}. The former trains $M$ distinct models simultaneously and estimates the uncertainty from their joint outputs. The latter applies spectral normalization to all layers, providing a bi-Lipschitz guarantee on all intermediate representations, and uses a Gaussian process output layer to efficiently estimate the uncertainty. Since ensembling increases the training and inference cost by a linear factor $M$, \sysname uses the SNGP approach due to its lower overhead. 


%% file: sections/7_experiments.tex
\section{Experiments}
\label{sec:experiments}

\rewrite{Our evaluation of \sysname seeks to demonstrate that it robustly achieves state-of-the-art execution latency speedups on parameterized query workloads. We summarize our main results as follows:
\begin{itemize}
    \item An end-to-end implementation of \sysname on PostgreSQL substantially outperforms the built-in optimizer and Bao. Both RCE and ML models play large roles in achieving this speedup. (Section~\ref{subsec:exp_e2e})
    \item RCE discovers significantly better plans than existing candidate generation baselines. We also observe that RCE plans are frequently superior to exact-cardinality plans. (Section~\ref{subsec:analyze_rce})
    \item Using SNGP models is crucial to capturing speedups generated by RCE while minimizing query regressions. (Section~\ref{subsec:model_exp}).
    \item We release a dataset consisting of $\sim$14.2 years of query executions as a benchmark for future research in modeling approaches (Section~\ref{subsec:dataset_contribution}).
\end{itemize}}




\paragraph{Objectives.} \rewrite{To evaluate our methods, we use both RCE speedup $S^{\texttt{opt}}(RCE)$ (shortened as $\sopt$) and model speedup $\sm$, defined in Equations \ref{eq:oracle_speedup} and \ref{eq:model_speedup} respectively. We note that $\sm = p \cdot \sopt$, where $0\leq p \leq 1$ corresponds to the proportion of the speedup the model captures. 
Since the model may predict worse plans than the built-in optimizer, we also measure the query regression frequency $\freg$, defined as the proportion of test query instances the model does at least 10\% worse than the default optimizer on. 
The primary metrics for each of our components are:
\begin{enumerate}
    \item \textbf{End-to-end performance:} $\sm$
    \item \textbf{Candidate generation performance:} $\sopt$
    \item \textbf{Model performance:} $p$, $\freg$
\end{enumerate}}
\rewrite{\sysname aims to maximize $\sm$ by maximizing $p$ and $\sopt$, while minimizng $\freg$. We also report the 99th-percentile tail latency speedup, which may be relevant in applied scenarios. We define this as $P_{99}^{\texttt{method}}(W)=\frac{p_{99}(\{ExecTime(p_{\texttt{default}}, q) \forall \,q\in W\})}{p_{99}(\{ExecTime(p_{\texttt{method}}, q) \forall \,q\in W\})}$, where $p_{99}(C)$ denotes the 99th percentile highest value in a collection $C$.}

\subsection{Setup}
\label{subsec:setup}

\paragraph{Datasets and query extraction.}

We use two synthetic benchmarks: TPC-H (uniform and skewed with Zipf factor = 1, 10 GB ~\cite{tpchskewed}), and Stack, a database consisting of real-world StackExchange data~\cite{marcus2021bao}. 
TPC-H consists of 22 parameterized queries.
We use an augmented version of Stack with 87 parameterized queries: 42 from the original benchmark and 45 additional manually-written query templates.

All experiments were run using PostgreSQL 13.5 on Google Cloud Platform (GCP) n1-highmem-16 instances with 16 CPU cores, 108 GB of RAM, and 2 TB of SSD. Following~\cite{leis2015good}, we set shared\_buffers to 75 GB, effective\_cache\_size to 80 GB, and work\_mem to 4 GB to ensure that the entire dataset fits in memory. For TPC-H, we use the indexes defined in BenchBase~\cite{DifallahPCC13}. For Stack, we add indexes on all primary keys, foreign keys, and columns that appear in a predicate of any query. 


\paragraph{Query instance generation.} We follow the official TPC-H specification~\cite{tpch} to generate parameter values of each query template. 
For Stack, we synthetically generate parameter values so that every query instance returns nonempty results. 
This is accomplished by uniformly sampling rows from the result set of a derived query that selects column values for which parameterized predicates would produce at least one value. Range predicates are constructed by first sampling a single value in the manner, then sampling lower/upper bounds around this value. 

\paragraph{Training query execution.} For each query instance and plan hints, we execute the resulting plan three times to simulate a warm-cache scenario, and take the minimum latency as the ground truth. 
For slow queries, we executed each up to 8 times in parallel on the same machine, and observed negligible differences with the serial execution setting. 
We leave a full analysis of different execution scenarios to future work.

\paragraph{RCE hyperparameters.} Unless stated otherwise, we use the same values for all RCE hyperparameters in all of our experiments, demonstrating its efficacy even when untuned for specific benchmarks. 
We set the number of generations $G$ to 3, the exponent base $b$ to 10, the exponent range $m$ to 2, the number of perturbations per plan $N$ to 20, and the number of samples extracted from each generation $S$ to 20.
For each query template, we run RCE on the first 50000 query instances for Stack, and all query instances for TPC-H. 

\paragraph{Model details.} All of our experiments use a fixed base neural network with three layers of 64 hidden units each. We use Adam with learning rate 3e-4, ReLU activation functions, and 10-dimensional string embeddings. For SNGP models, we additionally apply spectral normalization to all dense layers, and replace the output dense layer with a random Fourier feature Gaussian Process with 128 random features. For all models, we fall back to the default plan if the predicted confidence is less than 0.9. For all queries, we use a 80/20 train/test split, and report results (speedups, regressions) on the test workload. We did not attempt to tune our models or perform model selection, although it is straightforward to do so by reserving a validation set from the training dataset. 

\begin{figure}[t!]
    \centering

    \begin{subfigure}{0.25\columnwidth}
        \centering
         \begin{tabular}{ll}
         \hline
         $\sm$        & \% of queries \\ \hline
         \textgreater{}1.2 & 64.4\%        \\ 
         \textgreater{}2x   & 32.2\%        \\
         \textgreater{}10x  & 14.9\%        \\
         \textgreater{}20x  & 4.6\%         \\ \hline
         \end{tabular}
         \caption{\rewrite{Summary of PostgreSQL \sysname speedups on Stack.}}
         \label{tab:e2e_stack}             
    \end{subfigure}    
    \hspace{1pt}
    ~
    \begin{subfigure}{0.36\columnwidth}
        \centering
        \includegraphics[width=\linewidth]{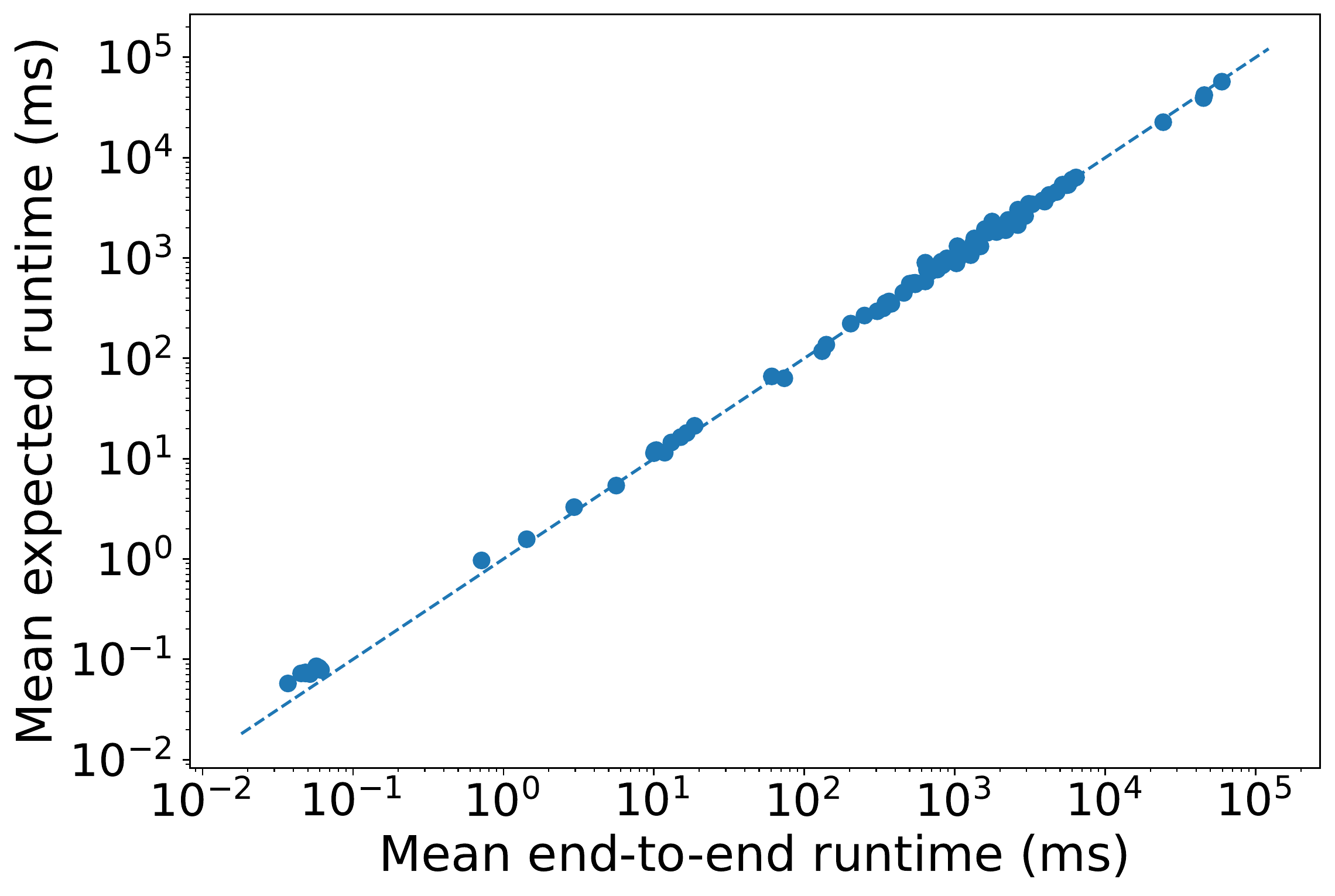}
        \caption{Actual end-to-end vs predicted latencies (ms).}
        \label{fig:e2e_stack}
    \end{subfigure}
    \hspace{1pt}
    ~
    \begin{subfigure}{0.36\columnwidth}
        \centering
        \includegraphics[width=\linewidth]{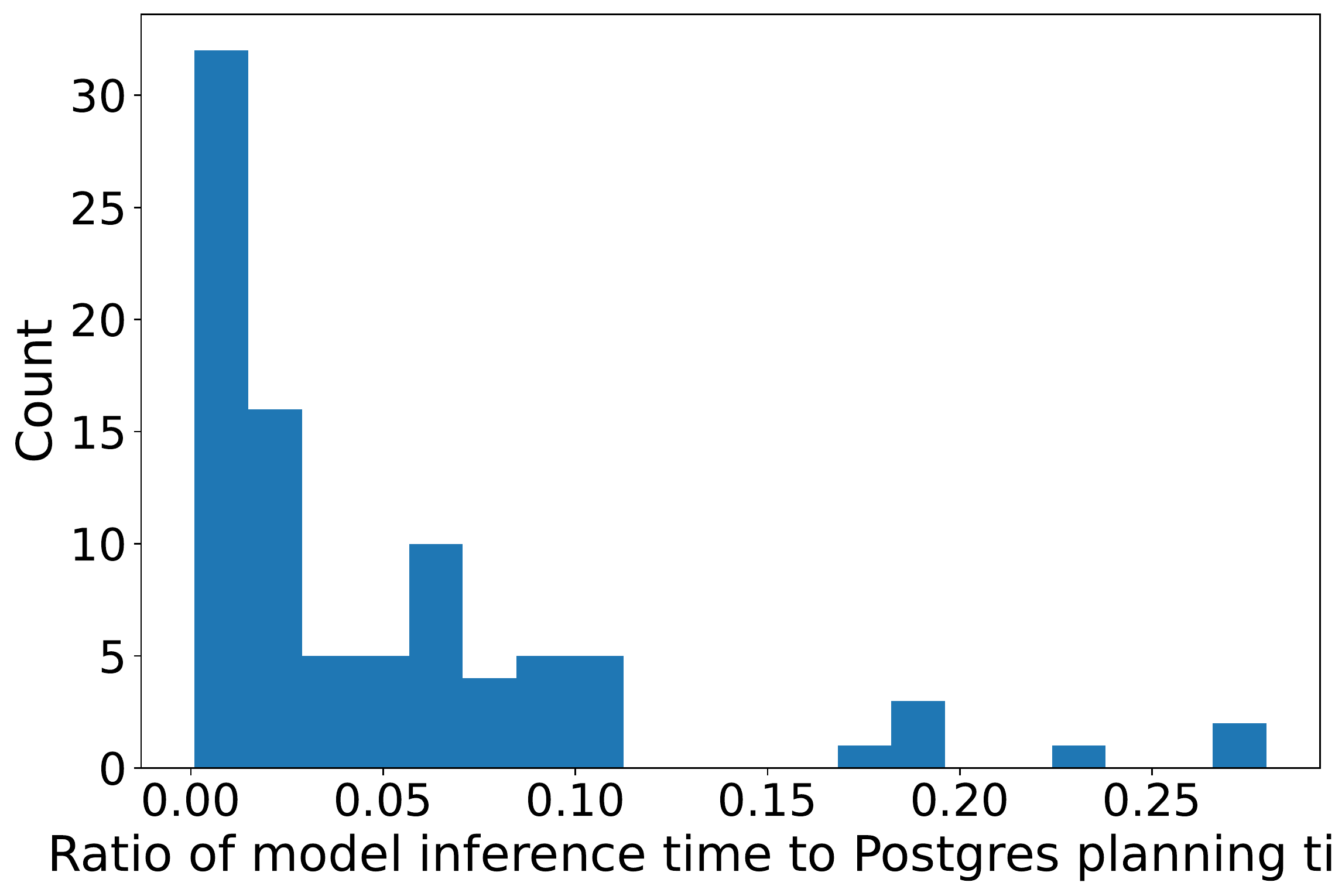}
        \caption{Histogram of model inference to PostgreSQL planning time ratios.}
        \label{fig:model_inference_histogram}
    \end{subfigure}
    
    \caption{PostgreSQL integration evaluation.}
    \label{fig:pg_e2e-eval}
    \vspace{-1em}
\end{figure}

\subsection{Kepler Improves Query Execution Latency}
\label{subsec:exp_e2e}

\begin{figure*}[ht!]
    \captionsetup{aboveskip=-1pt}
    \centering
    \includegraphics[width=\linewidth]{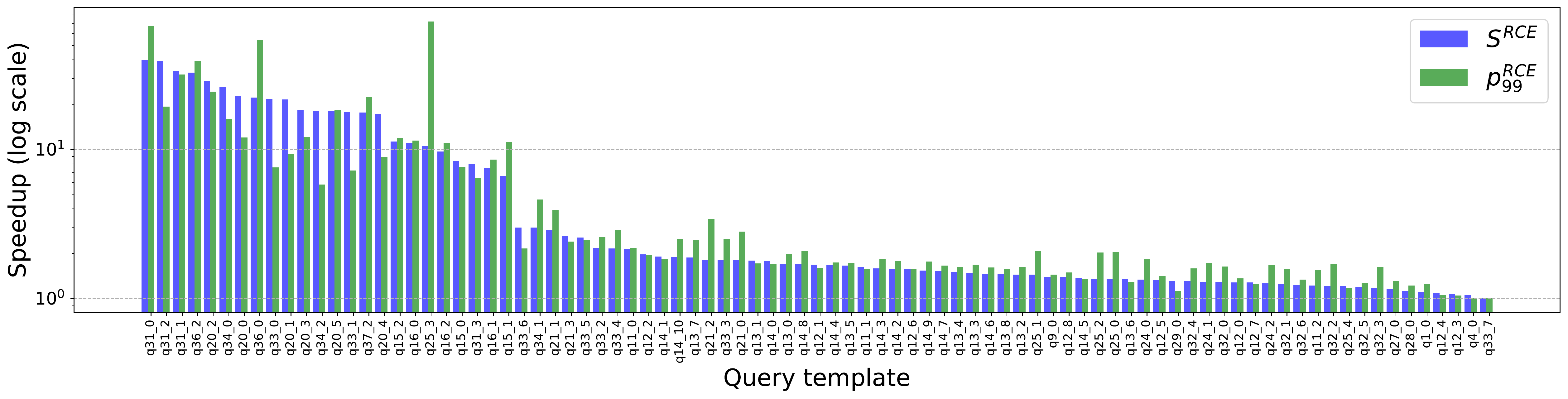}
    \caption{$\sopt$ and $\popt$ on Stack by query, sorted by $\sopt$}
    \label{fig:stack_headroom_p99}
\end{figure*}

\begin{figure}[t!]
\centering
\begin{subfigure}{0.49\columnwidth}
    \centering
    \includegraphics[width=\linewidth]{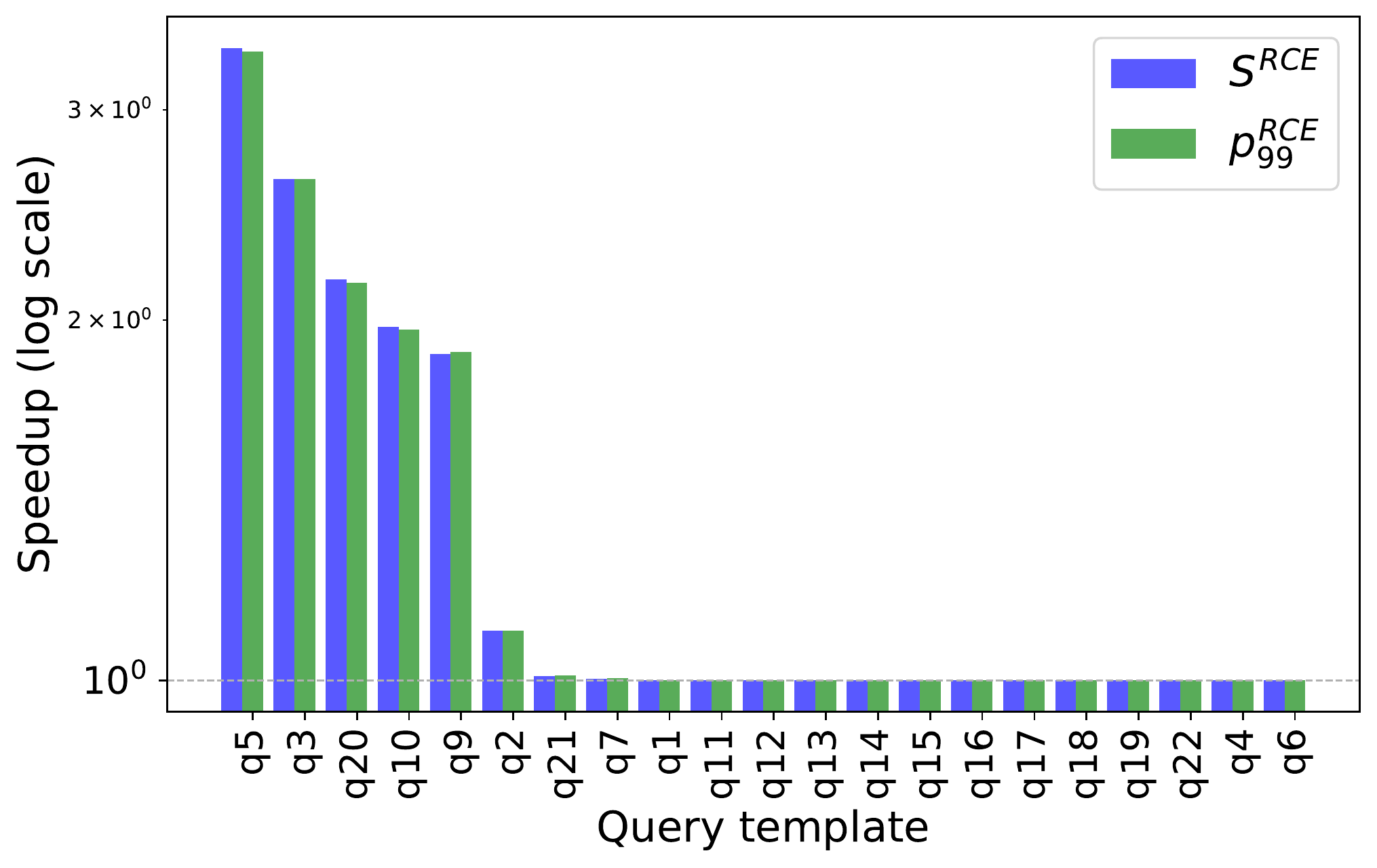}
    \caption{TPC-H.}
    \label{sub:tpch_headroom}
\end{subfigure}
~
\begin{subfigure}{0.49\columnwidth}
    \centering
    \includegraphics[width=\linewidth]{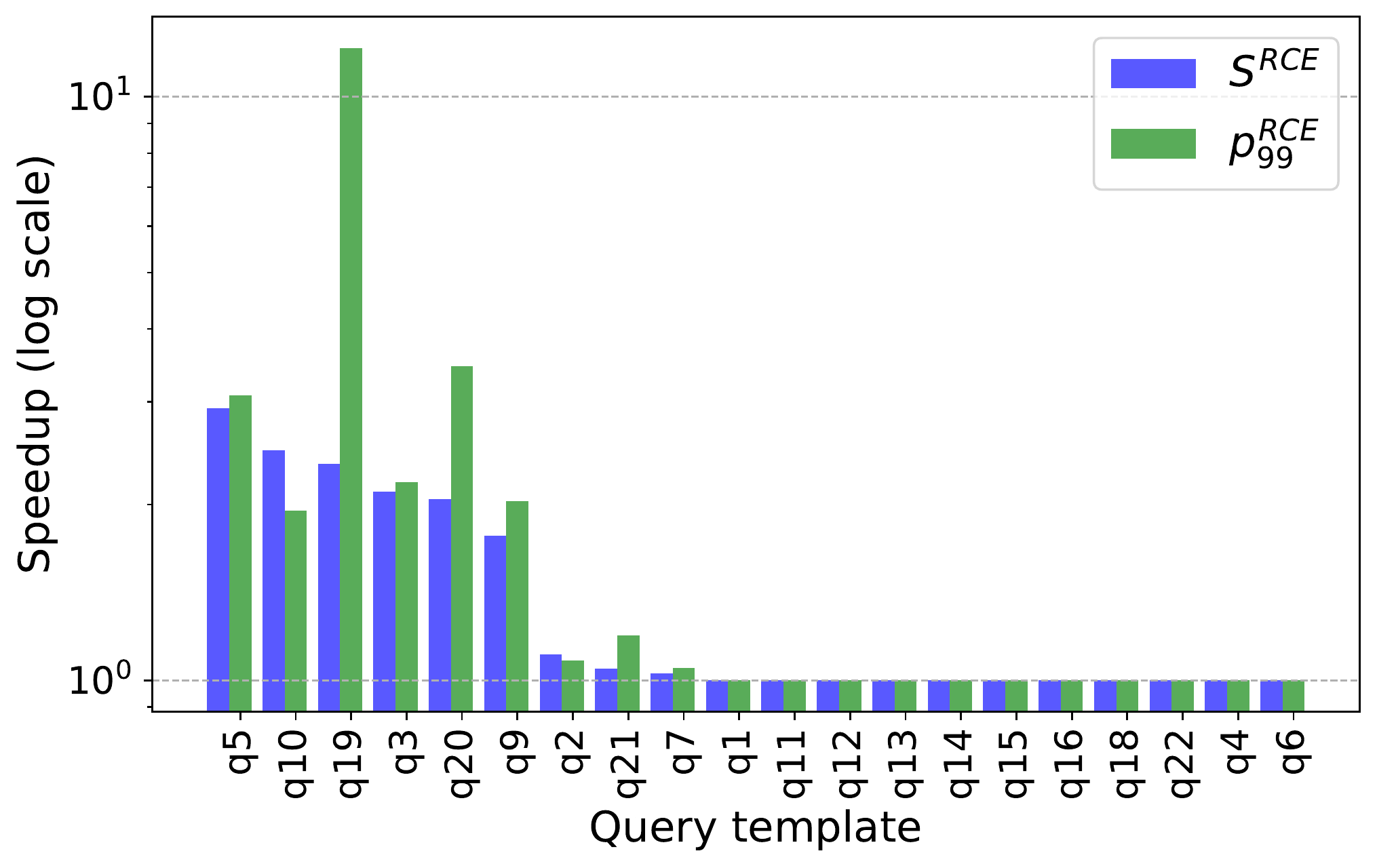}
    \caption{TPC-H skewed (z=1).}
    \label{sub:tpch_z1_headroom}
\end{subfigure}    
\caption{$\sopt$ and $\popt$ on TPC-H by query, sorted by $\sopt$.}
\end{figure}

\paragraph{End-to-end performance.} \rewrite{We integrate \sysname into the PostgreSQL query optimizer to demonstrate that it  delivers large speedups in a real deployment. Our implementation loads Tensorflow Lite models on the database server for fast CPU model inference and uses the $\texttt{pg\_hint\_plan}$ extension to force specific plans via hints. Providing the query id as a comment with the SQL query text from any PostgresSQL client connection triggers Kepler query plan prediction.} 

We executed a sample of 1000 evaluation set query instances per query on the integrated \sysname PostgresSQL system. Table~\ref{tab:e2e_stack} summarizes the speedups of \sysname over Stack, demonstrating that our PostgreSQL implementation achieves nontrivial speedups on the majority of queries, with over 2x speedup over the entire workload for 32.2\% of queries. These speedups indicate that \sysname is able to bypass inaccuracies PostgreSQL's cardinality estimation and cost model via RCE. 

 Figure~\ref{fig:e2e_stack} confirms that the \sysname deployment achieves near-identical speedups to those expected based on the pre-collected execution dataset. This is because the use of lightweight ML models and planning hints incur low planning-time overhead. Figure~\ref{fig:model_inference_histogram} shows the distribution of the ratio of model inference times to PostgreSQL planning time for all queries in Stack. The model inference time is mostly under 5\% of PostgreSQL planning time and at most 30\%.

Our total speedup results over entire workloads are quite significant since our workloads -- query instances sampled uniformly from the space of non-empty query instances -- are not designed to adversarially challenge the optimizer. Next, we summarize the contributions from the two key components: (1) RCE to uncover the potential speedups and (2) the ML models to capture speedups. Finally, we compare the results to Bao as a baseline.

\paragraph{RCE speedups.} 
\rewrite{We illustrate the efficacy of RCE by showing that it achieves large speedups on both Stack and TPC-H. Figure~\ref{fig:stack_headroom_p99} shows the per-template $\sopt$ and $\popt$, with RCE achieving over 2x speedup on 32/87 queries and over 1.2x speedup on 78/87 queries.}

\rewrite{Similarly, RCE improves 6/22 queries on TPC-H uniform (Figure~\ref{sub:tpch_headroom}) and 9/22 queries on TPC-H skewed (Figure~\ref{sub:tpch_z1_headroom}). In particular, RCE finds larger speedups on TPC-H skewed due to the non-uniformity in its data distribution.}

\begin{figure*}[!t]
    \centering
    \begin{subfigure}{0.45\linewidth}
        \centering
        \includegraphics[width=\linewidth]{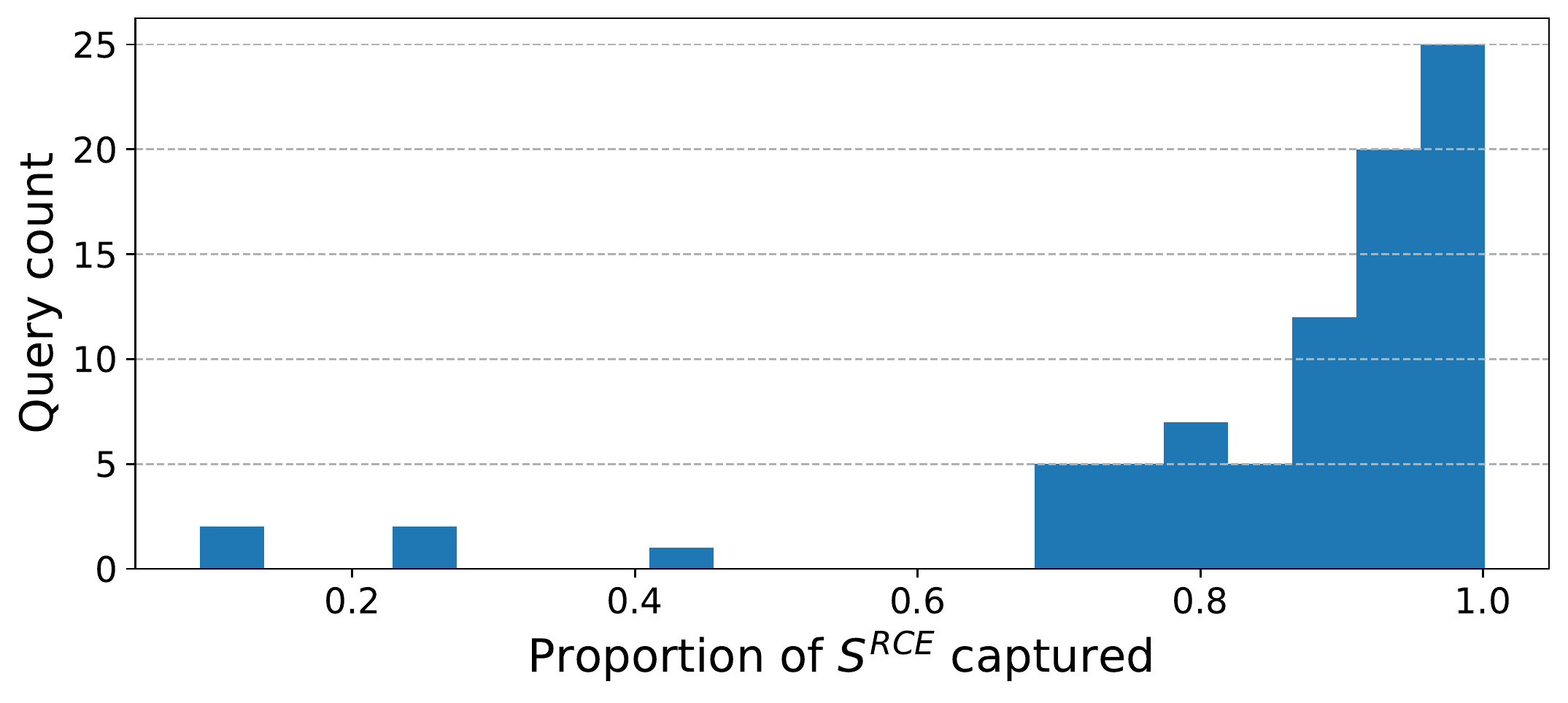}
        \caption{Stack, model speedup ratio histogram.}
        \label{fig:stack_model_speedups}
    \end{subfigure}
    \begin{subfigure}{0.45\linewidth}
        \centering
        \includegraphics[width=\linewidth]{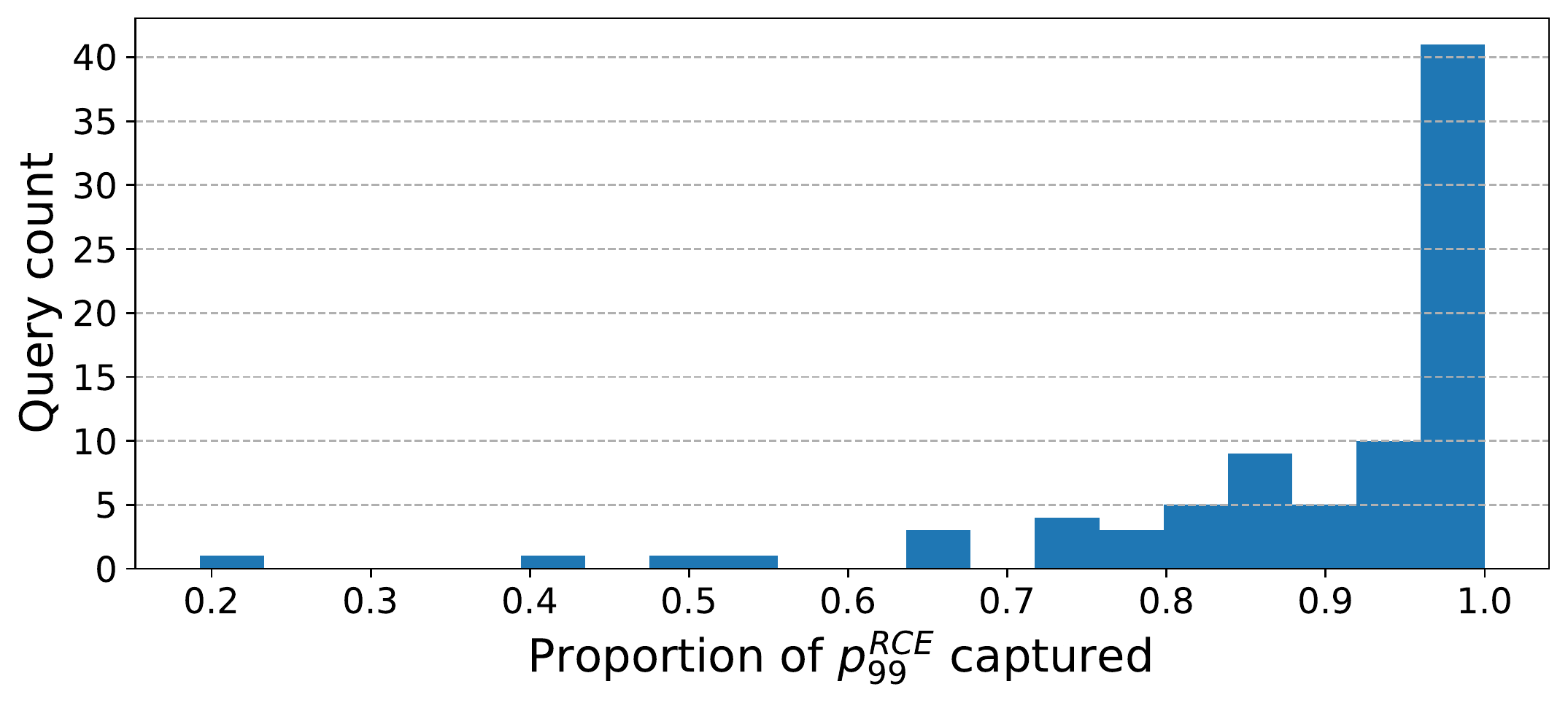}
        \caption{Stack, p99 model speedup ratio histogram.}
        \label{fig:stack_model_p99_speedups}
    \end{subfigure}
    \caption{Model results on Stack.}
\end{figure*}

\begin{figure*}[!t]
    \centering
    \includegraphics[width=0.7\linewidth]{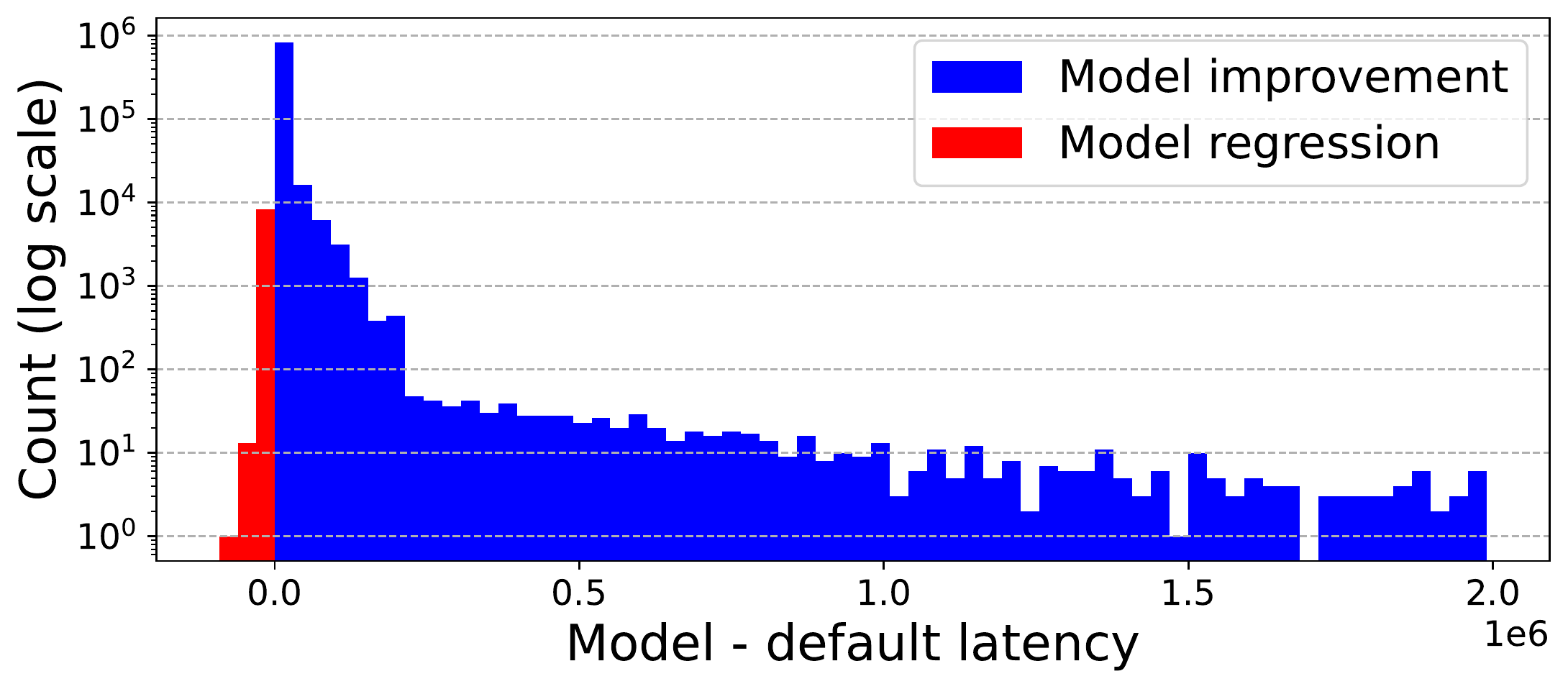}
    \caption{Query improvements and regressions on all query instances in Stack.}
    \label{fig:stack_model_regressions}
\end{figure*}

\paragraph{ML models predict fastest plans and avoid regressions.}

\rewrite{Next, we show that our ML models are able to robustly capture the speedup produced by RCE, i.e. they maximize $p$ while minimizing $\freg$. In Figures \ref{fig:stack_model_speedups} and \ref{fig:stack_model_p99_speedups}, we plot what proportion of $\sopt$ and $\popt$ on Stack we respectively capture per query. These distributions show that our models can reliably predict near-optimal plans: our models capture over 80\% of $\sopt$ on over 80\% of Stack queries. In Figure~\ref{fig:stack_model_regressions}, we plot the distribution of the absolute magnitude of model improvements and regressions compared to the default plan, illustrating that the frequency and magnitude of the regressions are minimal compared to those of the improvements.}

\paragraph{Bao on parameterized queries.} 
\begin{table}[t]
    \captionsetup{belowskip=-2pt}
    \caption{\rewrite{Comparison of Bao and \sysname performance.}}
    \begin{tabular}{lll}\hline
    Query  & Bao speedup & \sysname speedup \\\hline
    q11\_0 & 1.201                  & \textbf{1.651}             \\
    q12\_0 & 1.238                  & \textbf{1.271}             \\
    q13\_0 & 1.408                  & \textbf{1.566}             \\
    q14\_0 & \textbf{1.848}          & 1.619            \\
    q15\_0 & 1.157                  & \textbf{7.727}             \\
    q16\_0 & 1.168                  & \textbf{9.250}           \\ \hline
    \end{tabular}
    \label{tab:bao}  
\end{table}

We evaluate Bao, one of the few prior approaches that demonstrates actual improvement in execution latency, on our parameterized version of Stack~\cite{marcus2021bao}. \rewrite{For illustrative purposes, we run Bao for 2000 query instances on six representative templates from the original Stack dataset \cite{marcus2021bao}. As shown in Table~\ref{tab:bao}, we observed that \sysname outperforms Bao on 5 out of 6 templates, and in particular is able to find far greater speedups on q15\_0 and q16\_0. As we later show in Figure~\ref{fig:stack_bao_arms}, this is because the candidate generation algorithm in Bao is severely suboptimal.}

\paragraph{Training data collection cost.} \rewrite{The speedups achieved by \sysname come at a nontrivial training query execution cost - on average, we used 39 CPU days worth of query execution time per query template. Hence, \sysname is most applicable to workloads where query templates are executed at high frequency. We discuss directions to significantly reduce this training data collection cost at the end of Section~\ref{subsec:model_exp} as well as Sections \ref{subsec:dataset_contribution} and \ref{sec:future_work}.}

\subsection{Analyzing RCE}
\label{subsec:analyze_rce}

\begin{figure*}[t!]
    \centering
    \begin{subfigure}{0.6\linewidth}
        \centering
        \includegraphics[width=\linewidth]{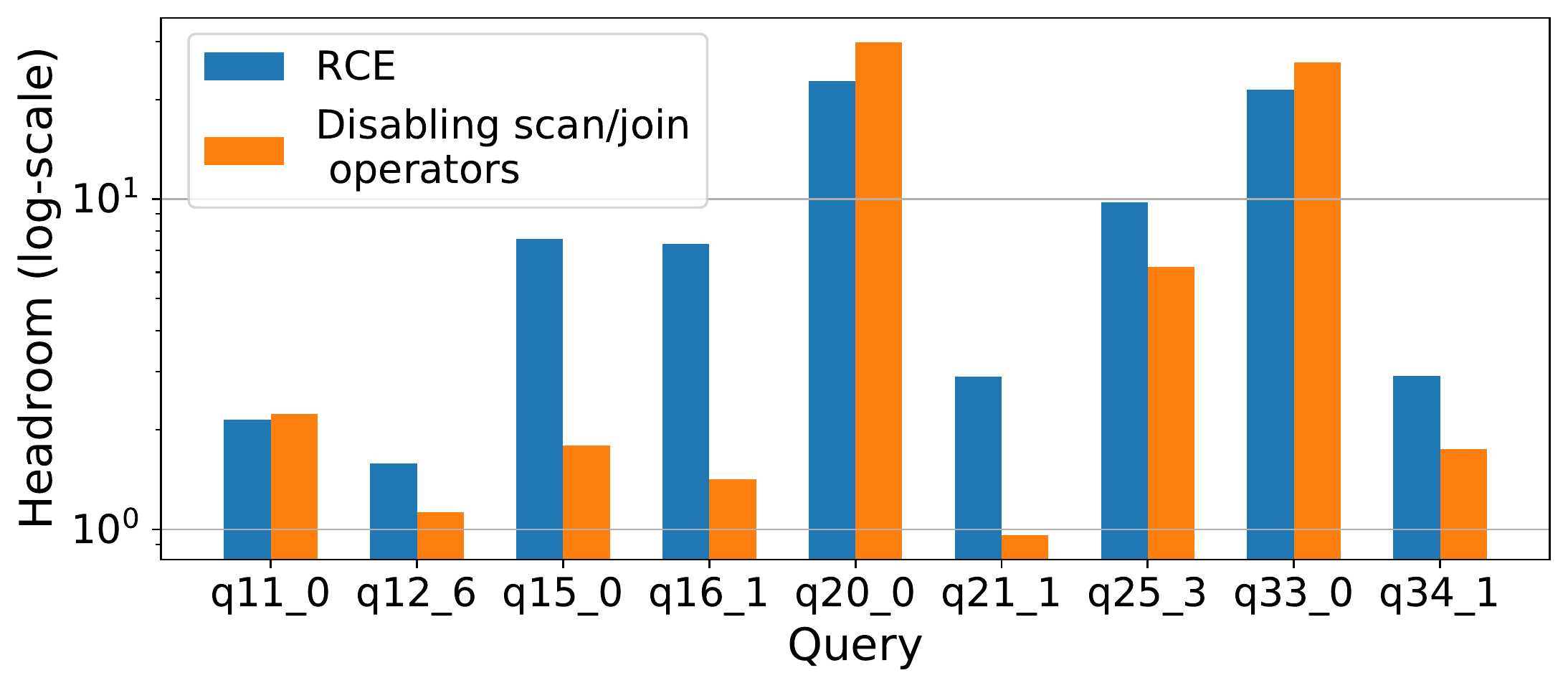}
        \caption{Comparison between RCE and Bao candidate generation (disabling scan/join methods).}
        \vspace{5mm}
        \label{fig:stack_bao_arms}
    \end{subfigure}
    \hspace{1pt}
    ~
    \begin{subfigure}{0.9\linewidth}
        \centering
        \begin{tabular}{lccc}
        \hline
        \toprule
        Algorithm                    & Speedup $S$       & Tail speedup $P99$ & \% improved \\
        \midrule
        RCE                      & 3.14          & 3.65        & 89.0\%                              \\ 
        RCE + PCP (ours)               & \textbf{3.11} & \textbf{3.63}        & \textbf{87.9\%}                     \\
        RCE + CBP           & 1.37          & 1.77        & 50.5\%                              \\ 
        PG                 & 2.0           & 2.07        & 49.4\%                              \\ 
        PG + CBP~\cite{vaidya2021leveraging} & 1.96          & 2.02        & 49.4\%                              \\
        \bottomrule
        \end{tabular}
        \caption{Comparison between PG and RCE with alternative pruning algorithms. We report average and 99th-percentile speedup aggregated using geometric mean, as well as the percentage of queries having total speedup greater than 20\%. PG + CBP corresponds to the method in \cite{vaidya2021leveraging}.}
        \label{tab:pruning}
    \end{subfigure}    
\caption{Candidate generation baseline comparisons on Stack.}
\end{figure*}

\rewrite{Having observed the behavior of the overall system, we now discuss characteristics of the first major component of \sysname: RCE. We evaluate RCE's performance against candidate generation and plan pruning baselines before exploring the effects of index configuration and key hyperparameter choices. After discussing RCE's empirical performance against those of exact cardinality (EC) plans, we close the section by offering perspectives and empirical justifications for why RCE works well.}



\paragraph{Comparison against baselines.} We compare against two main candidate generation baselines: 
\begin{itemize}
    \item \rewrite{PG: the default candidate generation method (Section~\ref{sec:related}) using the PostgreSQL optimizer.}
    \item \rewrite{Bao: for each query instance, we try every Bao arm, i.e. all 48 valid combinations on disabling various join/scan methods in PostgreSQL~\cite{marcus2018deep}.} 
\end{itemize}

\rewrite{Since \cite{vaidya2021leveraging} consider PG + cost-based pruning (CBP) as their candidate generation method, we simultaneously compare RCE against PG and our plan-cover pruning (PCP) algorithm against CBP in Table~\ref{tab:pruning}. RCE finds significantly more speedups than PG: $\sopt$ is over 1.2 on 89\% of Stack queries, as opposed to 49.4\% for PG. Although CBP achieves little speedup loss when applied to PG, it incurs far greater loss when applied to RCE, indicating that optimizer cost estimates cannot reliably predict the quality of RCE plans. By contrast, PCP achieves almost no degradation in speedup since it uses actual execution data to evaluate plans.}

\rewrite{The Bao candidate generation method produces far more plans than RCE, so for computational reasons we evaluated it on a diverse subset of queries designed to be representative of the entire Stack dataset.} Figure~\ref{fig:stack_bao_arms} shows that RCE achieves similar or better speedup on all queries, and is notably able to find non-trivial speedup on each query.


\begin{table}[t!]
\caption{\rewrite{Effect of number of generations in RCE in Stack query q11\_0.}}
\label{tab:stack_depth}
\begin{tabular}{llll}
\hline
$G$ & Total num plans & Plan cover size & $\sopt$ \\ \hline
1              & 9              & 6               & 1.369   \\
2              & 29             & 7               & 2.185   \\ 
3              & 62             & 9               & 2.235   \\ 
4              & 99             & 8               & 2.166   \\ \hline
\end{tabular}
\end{table}

\paragraph{Number of generations $G$.}\rewrite{The hyperparameter $G$ has a significant impact on the number and quality of RCE-generated plans, which we illustrate by varying $G$ for q11\_0 in Stack, with results shown in Table~\ref{tab:stack_depth}. The number of plans and plan cover size steadily grows up to three generations, although most of the speedup is captured in plans found using two generations. Increasing $G$ to four generations does not produce any marginal benefit.}


\paragraph{Exponential row count perturbation range: $b$ and $m$.} 
\rewrite{
To justify our choice of perturbation range hyperparameters $b=10$ and $m=2$, we analyzed the perturbation factor necessary to induce a change at a single level in the plan. For each of $n-1$ sub-plans in a query tree (for a query $Q$ with $n$ tables), we use binary search to identify the factor by which the cardinality estimate for that sub-plan must be perturbed in order to induce a change in the optimizer's plan. On Stack, we observed that although some plans needed a $10^6$ perturbation factor to induce a plan change, the vast majority of factors were less than $10^2$.
}

\begin{table}
\caption{Geometric means of per-query $\sopt$ for various index configurations.}
\label{tab:stack_index_compare}
\begin{tabular}{lllll}
\hline
          PK & +FK   & +Predicates & +DBA \\
         \hline
 1.888   & 2.41 & 5.286         & 5.284       \\ \hline       
\end{tabular}
\end{table}

\paragraph{Robustness to index configuration.}\rewrite{ RCE finds better plans on databases with different index configurations. We ran RCE over the following index configurations for Stack: primary keys only (PK), foreign keys (FK), predicate columns, and database administrator defined additional indexes~\cite{marcus2021bao}. Table~\ref{tab:stack_index_compare} shows RCE finds faster plans in all configurations. 
Similar to~\cite{leis2015good}, we find that more indexes leads to a larger speedup.}

\begin{table}
    \centering
    \caption{Comparison of average query latencies (in seconds) for RCE best plan, exact cardinality plan, and PostgreSQL selected plan.}
    \label{tab:rce_vs_exact}    
    
    \begin{tabular}{llll}
    \hline
    Query   & RCE            & Exact Cardinality & Query optimizer \\ \hline
    q11\_0  & \textbf{0.045} & 0.167             & 0.180           \\ 
    q12\_2  & \textbf{0.350} & 0.768             & 0.954           \\ 
    q14\_10 & 2.503          & \textbf{2.107}    & 5.732           \\
    q16\_1  & \textbf{1.136} & 2.174             & 8.575           \\ 
    q20\_0  & \textbf{0.012} & 0.046             & 0.232           \\
    q33\_0  & \textbf{0.519} & 2.920             & 3.377           \\ \hline
    \end{tabular}
\end{table}

\paragraph{Can RCE discover optimal plans?} Although RCE demonstrably generates faster plans than a variety of baselines, we would also like to know how close are RCE-generated plans to the true optimal plans $p^*_{\texttt{opt}}(q)$. Since it is infeasible to determine $p^*_{\texttt{opt}}(q)$ in practice, we instead compare against the standard benchmark of exact cardinality plans~\cite{leis2015good, negi2021flow}. 

\rewrite{
We executed exact cardinality plans for a subset of the training workload and compared them against their respective best RCE plans in Table~\ref{tab:rce_vs_exact}. We again used a subset of queries from the Stack dataset for computational reasons. Due to the large number of joins in some query templates, we additionally set the exact cardinality for a subset of tables to be a high constant if its corresponding query did not finish within 15 minutes. Notably, RCE plans are substantially faster than exact cardinality plans on 5 out of 6 queries. This illustrates that even accurate cardinality estimation methods can lead to suboptimal plans due to incorrect assumptions and other deficiencies in the cost model.}

\paragraph{Query clusters facilitate RCE.} 
\begin{table}
    \centering
    \caption{Comparison of average query latencies (in seconds) for RCE-all, RCE-cluster, and RCE-instance.}
    \label{tab:cluster}    
    
    \begin{tabular}{llll}
    \hline
    Query  & RCE-all     & RCE-cluster     & RCE-instance \\ \hline
    q11\_1 & 0.024       & 0.027           & 0.051                   \\
    q12\_6 & 0.307       & 0.307           & 0.316                   \\
    q20\_0 & 0.006       & 0.006           & 0.010                   \\
    q21\_1 & 0.129       & 0.129           & 0.153                   \\
    q25\_3 & 2.212       & 2.212           & 2.212                   \\
    q33\_0 & 0.593       & 0.670           & 0.810                   \\
    q34\_1 & 2.482       & 2.663           & 2.914                   \\ \hline
    \end{tabular}
\end{table}

Query instances with similar parameter binding values often have similar query plan behavior, e.g. as visualized by plan diagrams~\cite{haritsa2005analyzing}. We hypothesize that these groups of similar instances, or \emph{clusters}, can dramatically increase the efficacy of RCE via plan sharing. 
Recall that in our candidate generation procedure, we execute RCE for each query instance and take the union over all resulting plan sets. 
Hence, each cluster only requires a single query instance's RCE process to reach the cluster-wide optimal plan. 
For a cluster of size $N$ and probability $\mathbb{P}(q_i)$ of query instance $q_i$ discovering the optimal plan, the overall probability of discovering the plan over the cluster is $1 - \prod_{i=1}^N \left(1 - \mathbb{P}(q_i)\right)$, which rapidly approaches 1 for sufficiently large $N$ and a reasonable distribution of $\mathbb{P}(q_i)$.

\rewrite{To demonstrate the existence and impact of clusters, we define a cluster for each plan $p$ as all query instances for which $p$ is the fastest plan.} Then, for each query instance $q$, we compare the execution latencies of the fastest plan from three plan sets: (1) RCE-all, containing all plans from all query instances, (2) RCE-cluster, containing all plans from query instances in the same cluster as $q$, and (3) RCE-instance, containing only the plans generated from the RCE process for $q$. As shown in Table~\ref{tab:cluster}, RCE-all and RCE-cluster have very similar execution times, while RCE-instance is often slower, indicating that intra-cluster plan sharing plays a large role in the efficacy of RCE.



\subsection{ML Models}
\label{subsec:model_exp}

\rewrite{We first motivate the use of ML models by demonstrating that Stack is highly \emph{parameter sensitive} -- i.e. different query instances have different optimal plans. We then justify modeling design choices with an ablation of using SNGP in our models and evaluation of varying confidence thresholds highlight the importance of incorporating robustness as a primary design component in \sysname. We conclude with extensive analyses around feature space selection, embedding vocabularies, and training data size.}

\paragraph{Parameter sensitivity.} 
\begin{figure}[t!]
    \captionsetup{belowskip=-5pt}

    \centering
    \begin{subfigure}{0.5\linewidth}
        \centering
        \includegraphics[width=\linewidth]{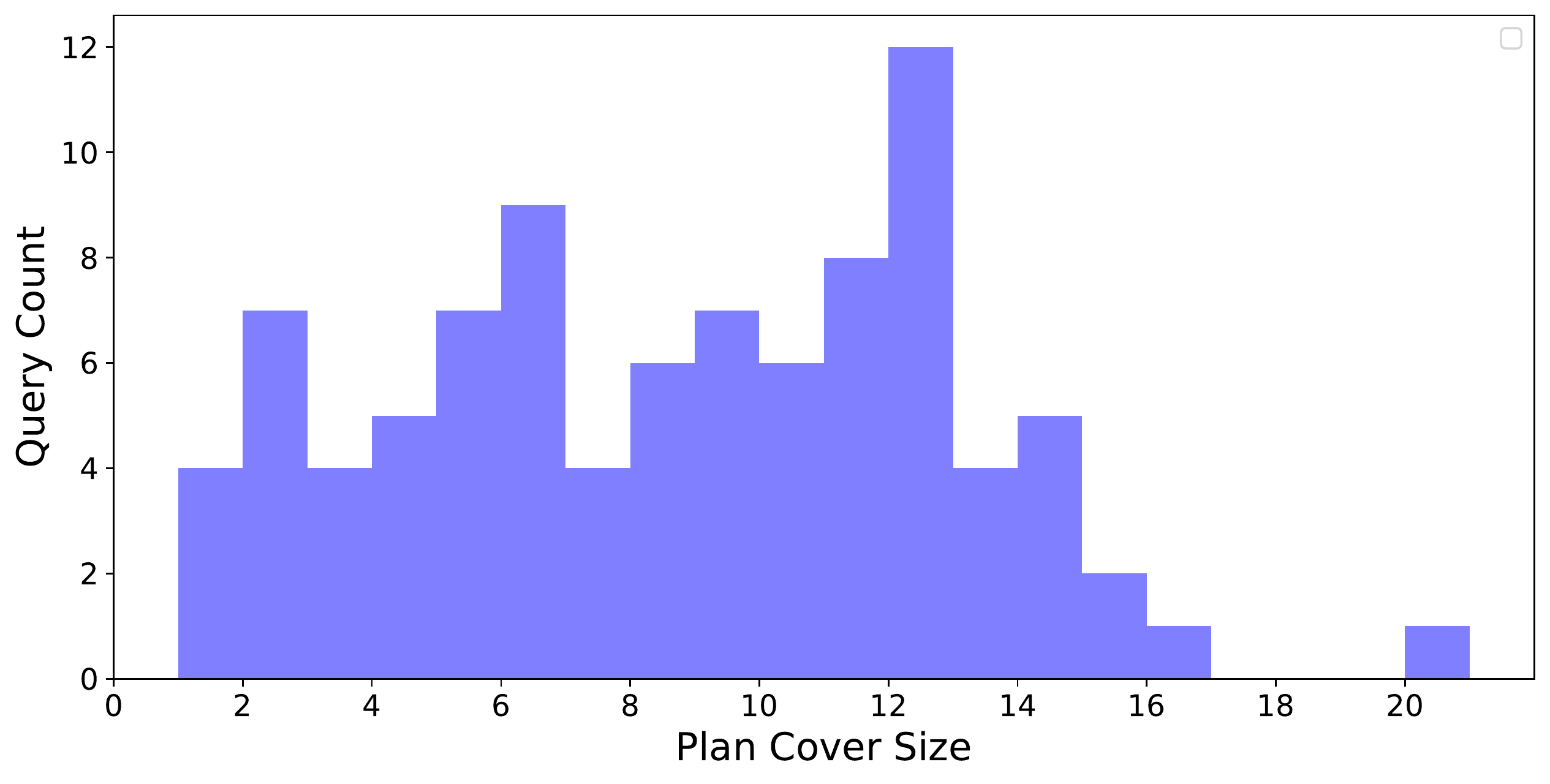}
        \caption{Histogram of plan cover sizes in Stack.}
        \label{fig:stack_plan_cover_sizes_histogram}
    \end{subfigure}
    ~
    \begin{subfigure}{0.5\linewidth}
        \centering
        \includegraphics[width=\linewidth]{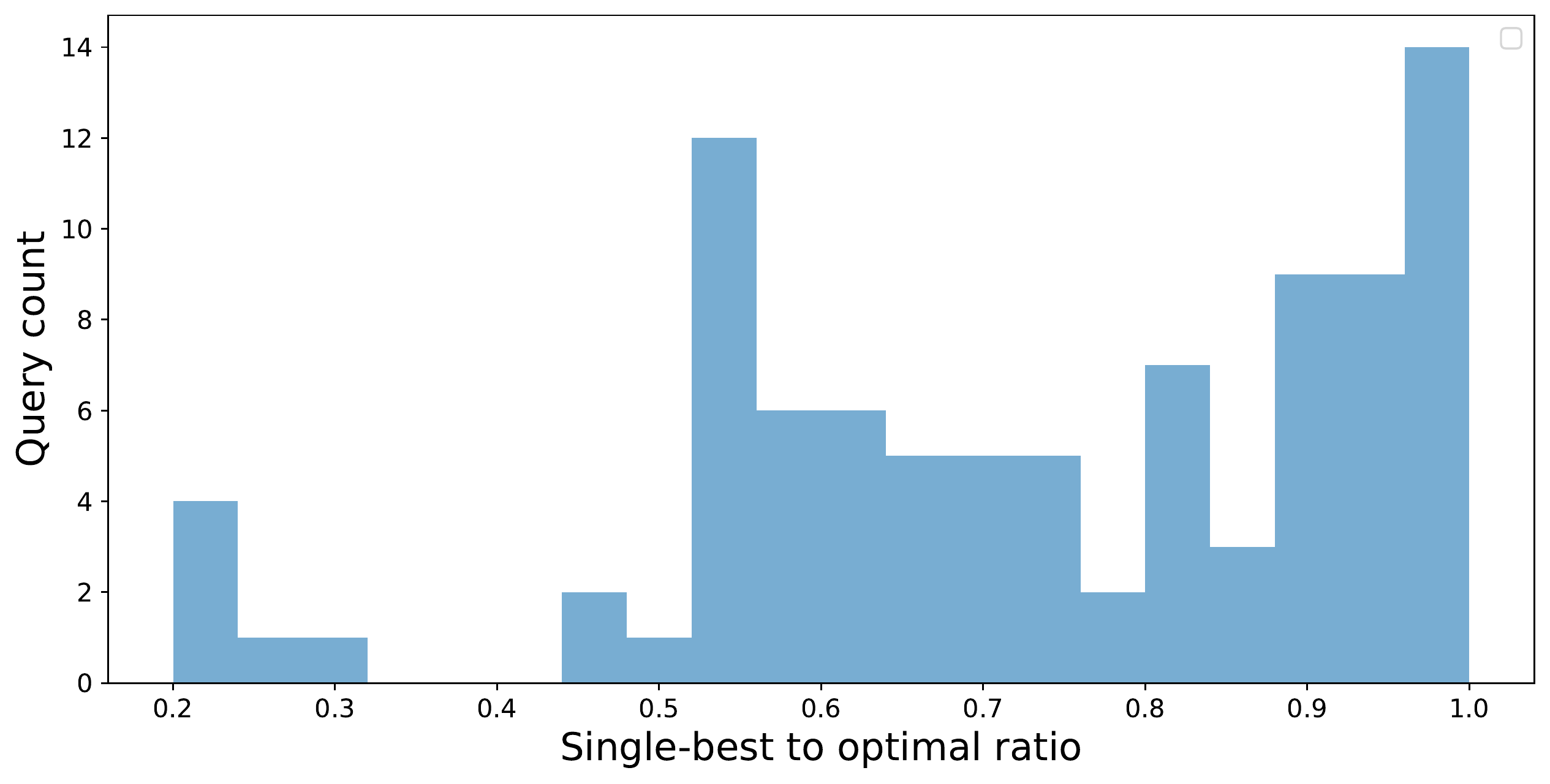}
        \caption{Histogram of single-best plan to all-plan $\sopt$ ratio.}
        \label{fig:stack_single_best_plan}
    \end{subfigure}
    \caption{Parameter sensitivity on all Stack queries.}
    \vspace{-1em}
\end{figure}

We investigate the parameter sensitivity of Stack and TPC-H queries based on our execution data. On Stack, all but four query templates had plan cover size greater than one (Figure~\ref{fig:stack_plan_cover_sizes_histogram}). \rewrite{In Figure~\ref{fig:stack_single_best_plan}, we show the distribution of \emph{single-best plan suboptimality ratios}, defined as the ratio of total latency of the oracle best plan against the total latency of the single best plan (i.e., the fixed plan with minimum total execution time). Ratios less than 1 indicate that using only a single plan incurs a loss in speedup, with lower values being more severely suboptimal. Thus, Figure~\ref{fig:stack_single_best_plan} implies that multiple plans are necessary to capture the full speedup.}

On the other hand, TPC-H is designed to not be parameter sensitive, which we confirmed by observing a plan cover of size 1 for all queries. Hence, our modelling results focus solely on Stack.

\paragraph{Loss functions/training objectives.} 

\begin{figure}[t!]
    \centering
    \begin{subfigure}{0.49\columnwidth}
        \centering
        \includegraphics[width=\linewidth]{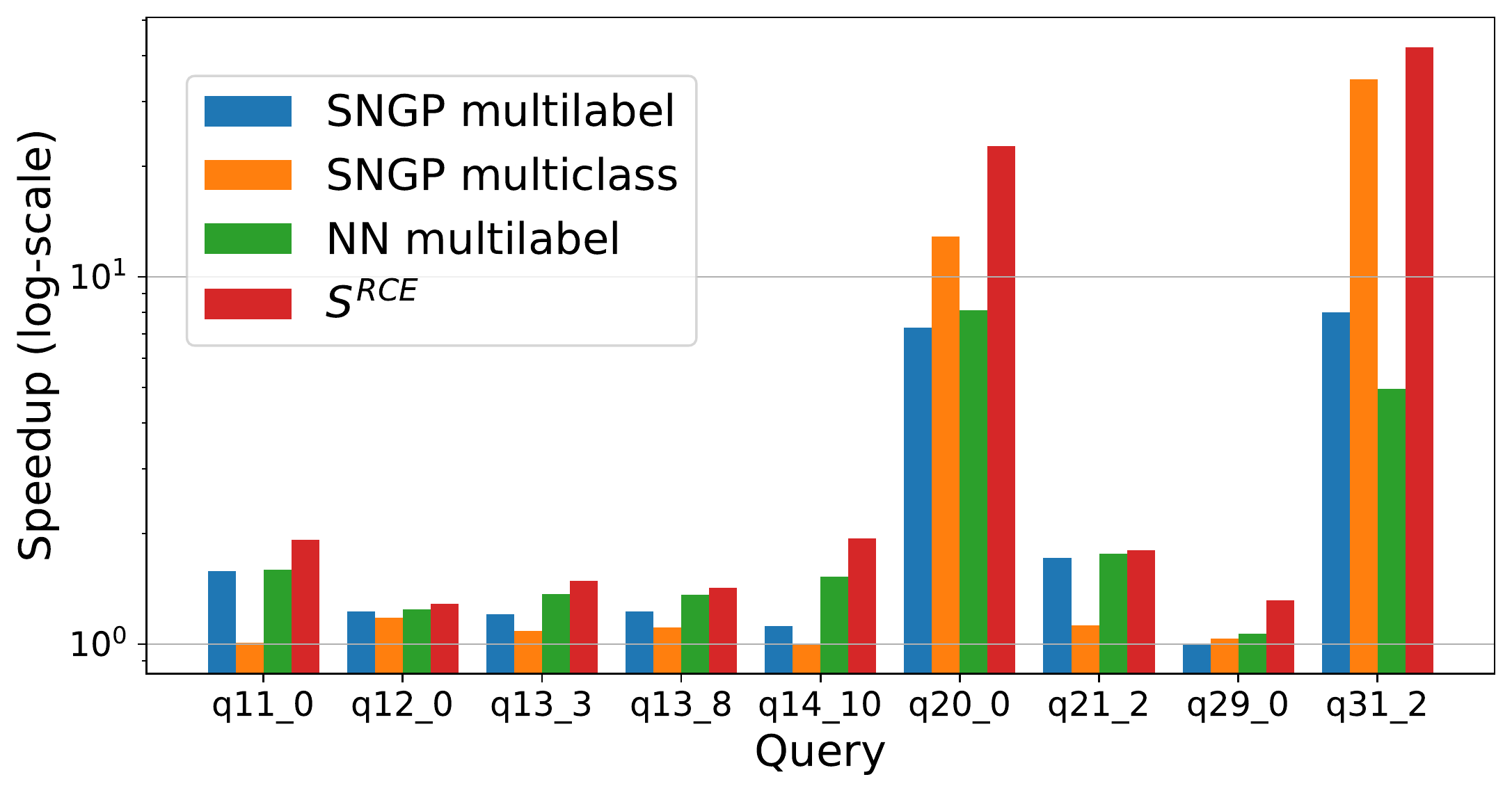}
        \caption{$\sm$}
        \label{sub:sngp_speedup}
    \end{subfigure}
    ~
    \begin{subfigure}{0.49\columnwidth}
        \centering
        \includegraphics[width=\linewidth]{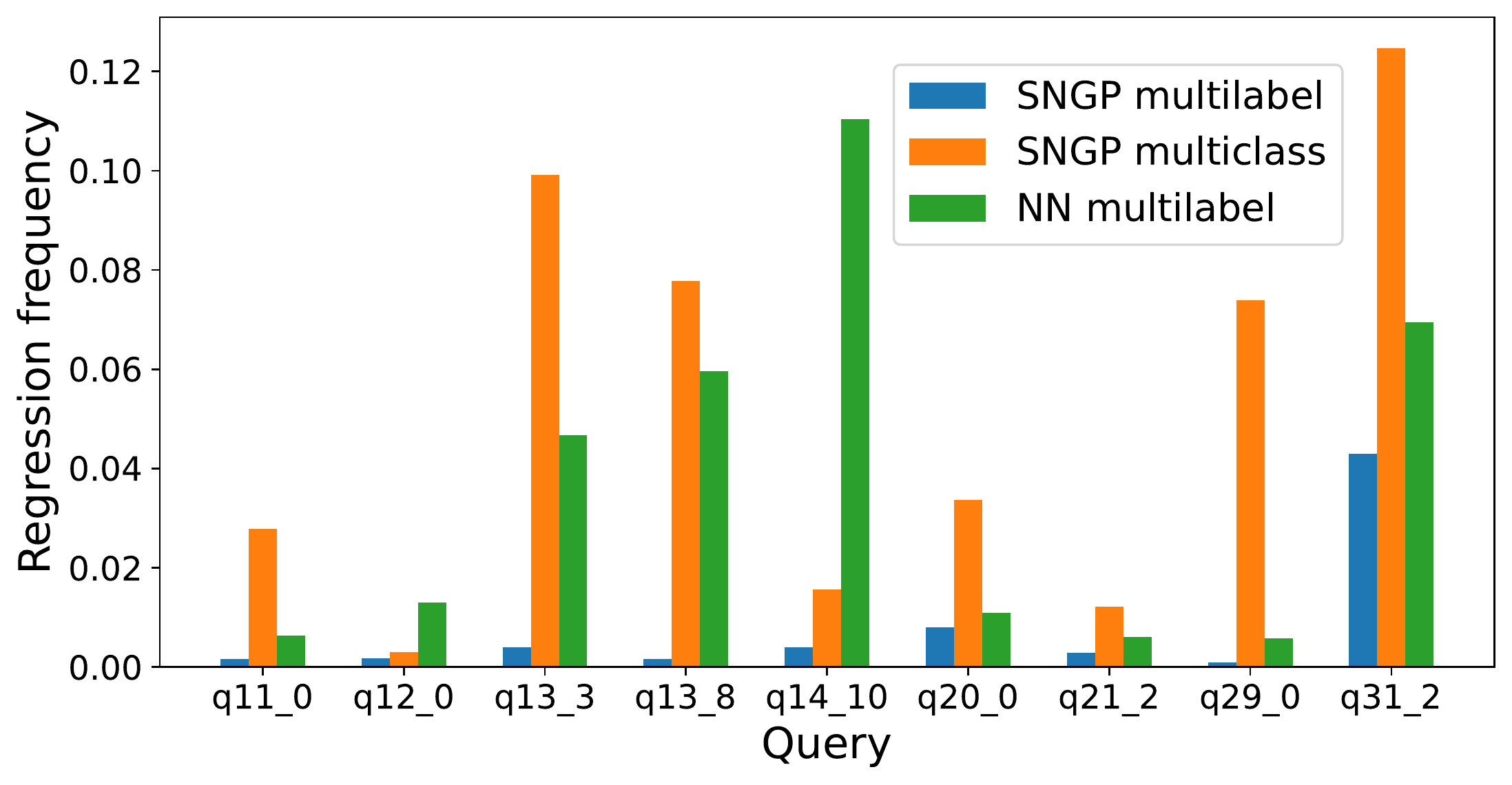}
        \caption{$\freg$}
        \label{sub:sngp_regression}
    \end{subfigure}    
    \caption{Comparison of uncertainty approaches.}
    \label{fig:loss_comparison}
\end{figure}

Figure~\ref{fig:loss_comparison} compares various loss functions and models: SNGP with multilabel loss, SNGP with multiclass loss, and a vanilla NN with multilabel loss. Multilabel loss + SNGP achieves similar or better speedup to other methods, while having a lower regression frequency for all queries.

\paragraph{Model calibration and uncertainty.} 
\begin{figure}[!t]
    \centering
    \includegraphics[width=0.8\linewidth]{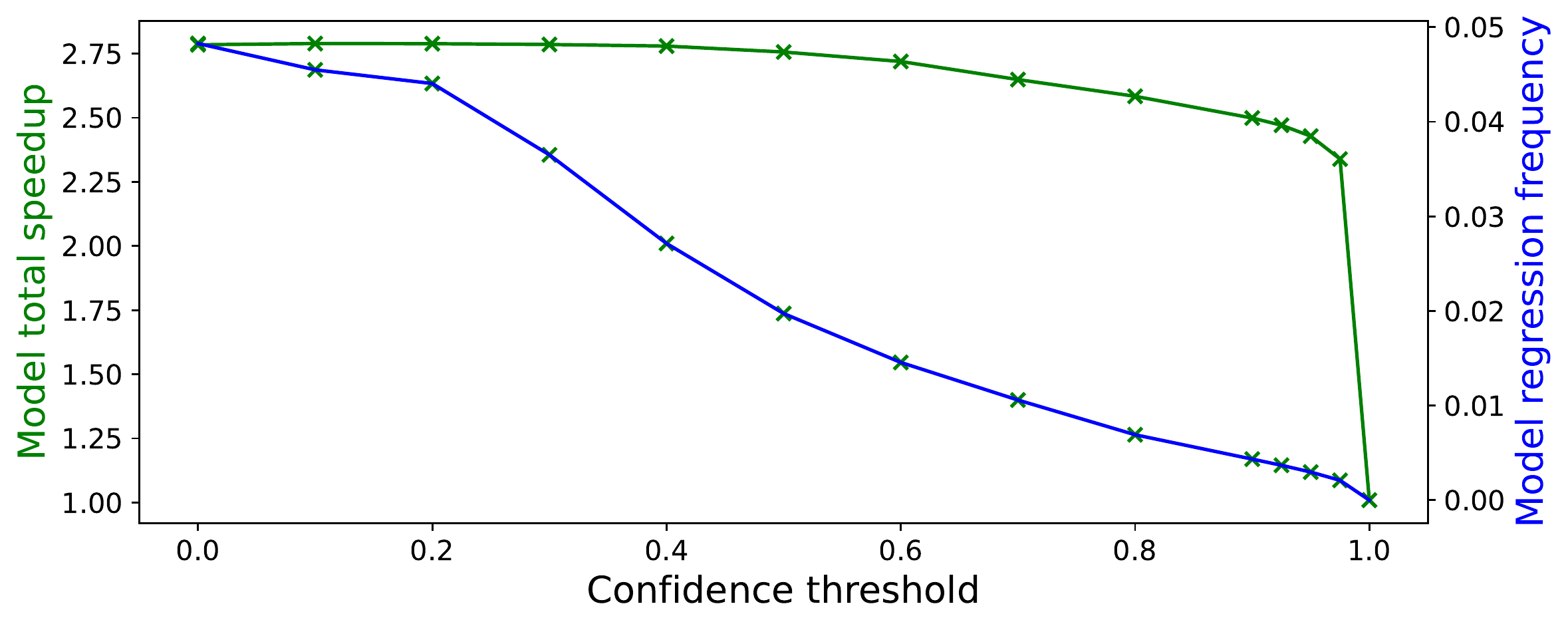}
    \caption{Tradeoff between $\sm$ and $\freg$ (aggregated respectively using geo-mean and mean over all Stack queries.}
    \label{fig:pareto_confidence}
\end{figure}

We further investigate the ability of our SNGP models in producing calibrated output probabilities. In Figure~\ref{fig:pareto_confidence}, we plot the test workload model speedup and regression frequency as a function of confidence threshold varying from 0 (no falling back) to 1 (always falling back). The captured speedup and regression frequency both decay smoothly as a function of confidence, which allows the user to specify their tolerance for regressions. This figure also demonstrates the importance of the fallback mechanism: one can dramatically reduce the regression frequency while only sacrificing a small portion of the speedup.

\begin{figure}[t!]
    \centering
    \begin{subfigure}{0.49\columnwidth}
    \resizebox{\textwidth}{!}{
        \begin{tabular}{llll}
        \hline
              & $\sopt$ & $\sm$ & $\freg$ \\ \hline
        No fallback & 1.044          & 0.781         & 23.8\%               \\ \hline
        Fallback  & 1.044          & 0.997         & 0.2\%                \\ \hline
        \end{tabular}}
        \caption{Site name.}
        \label{fig:site_ood}
    \end{subfigure}
    \hfill
    \begin{subfigure}{0.49\columnwidth}
    \resizebox{\textwidth}{!}{
        \begin{tabular}{llll}
        \hline
              & $\sopt$ & $\sm$ &  $\freg$ \\ \hline
        No fallback & 2.027          & 1.866         & 3.3\%                \\ \hline
        Fallback  & 2.027          & 1.822         & 0.1\%                \\ \hline
        \end{tabular}}
        \caption{Question last activity date.}
        \label{fig:date_ood}
    \end{subfigure}    
    \caption{\rewrite{Out-of-distribution experiments.}}
    \label{fig:ood}
\end{figure}

\paragraph{SNGP out-of-distribution detection.} \rewrite{Although \sysname is designed for relatively static workloads, it is robust to dynamic workloads by falling back to the default plan for out-of-distribution (OOD) inputs. We evaluate SNGP's OOD detection ability by holding out specific slices of the training distribution for an example query, q21\_2 on Stack. We consider two variants: (1) holding out sites totaling up to 20\% of the workload, and (2) holding out the last 20\% of last\_activity\_date values on the question table. Figure~\ref{fig:ood} shows that in both scenarios, \sysname's fallback mechanism allows it accurately detect OOD inputs and drastically reduce $\freg$ while still preserving some speedup.}

\paragraph{Raw parameter values vs selectivity features.} 

\begin{figure}[t!]
    \centering
    \begin{subfigure}{0.49\columnwidth}
        \centering
        \includegraphics[width=\linewidth]{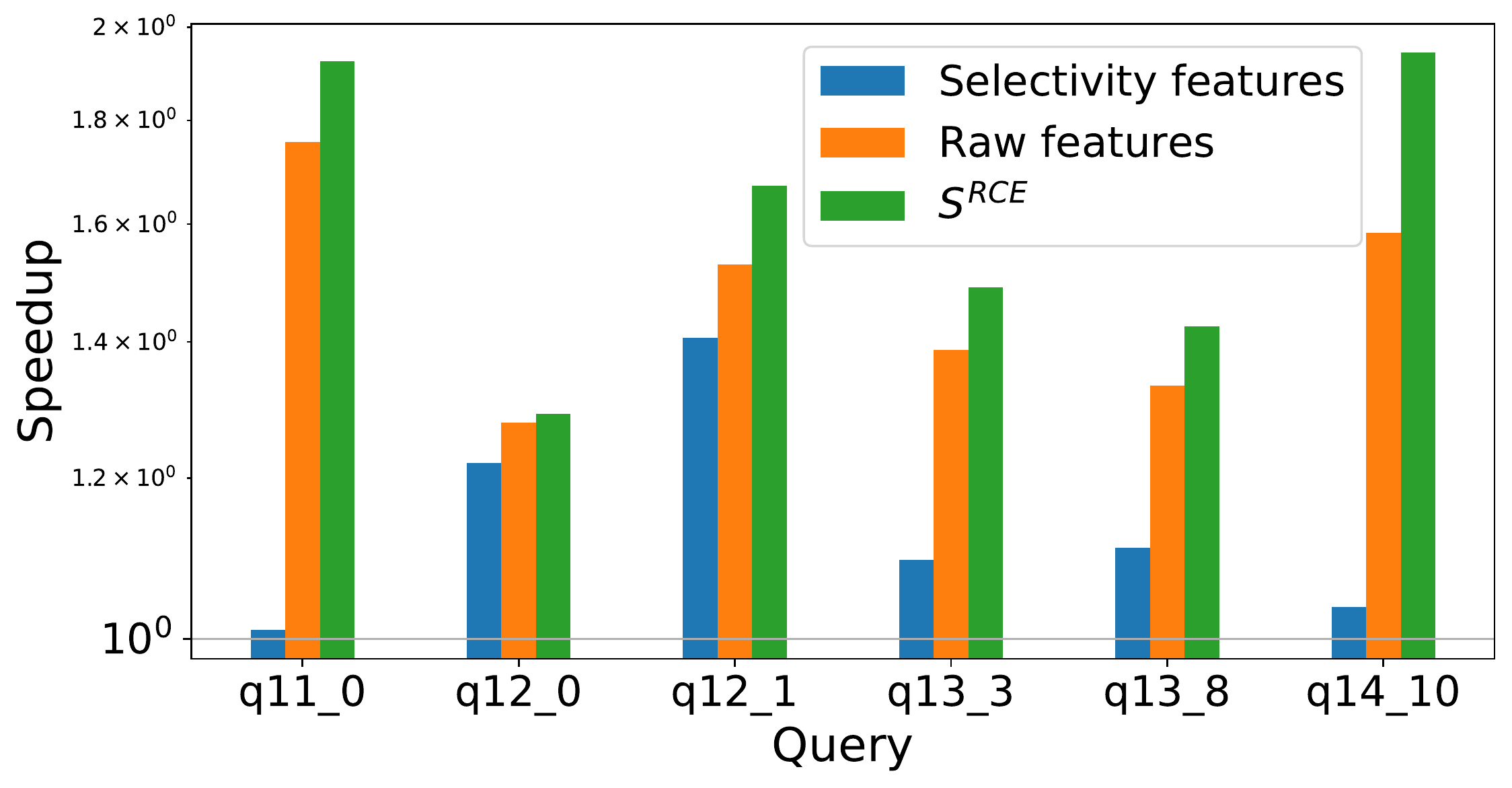}
        \caption{Selectivity vs raw features.}
        \label{fig:selectivity_baseline}
    \end{subfigure}
    ~
    \begin{subfigure}{0.49\columnwidth}
        \centering
        \includegraphics[width=\linewidth]{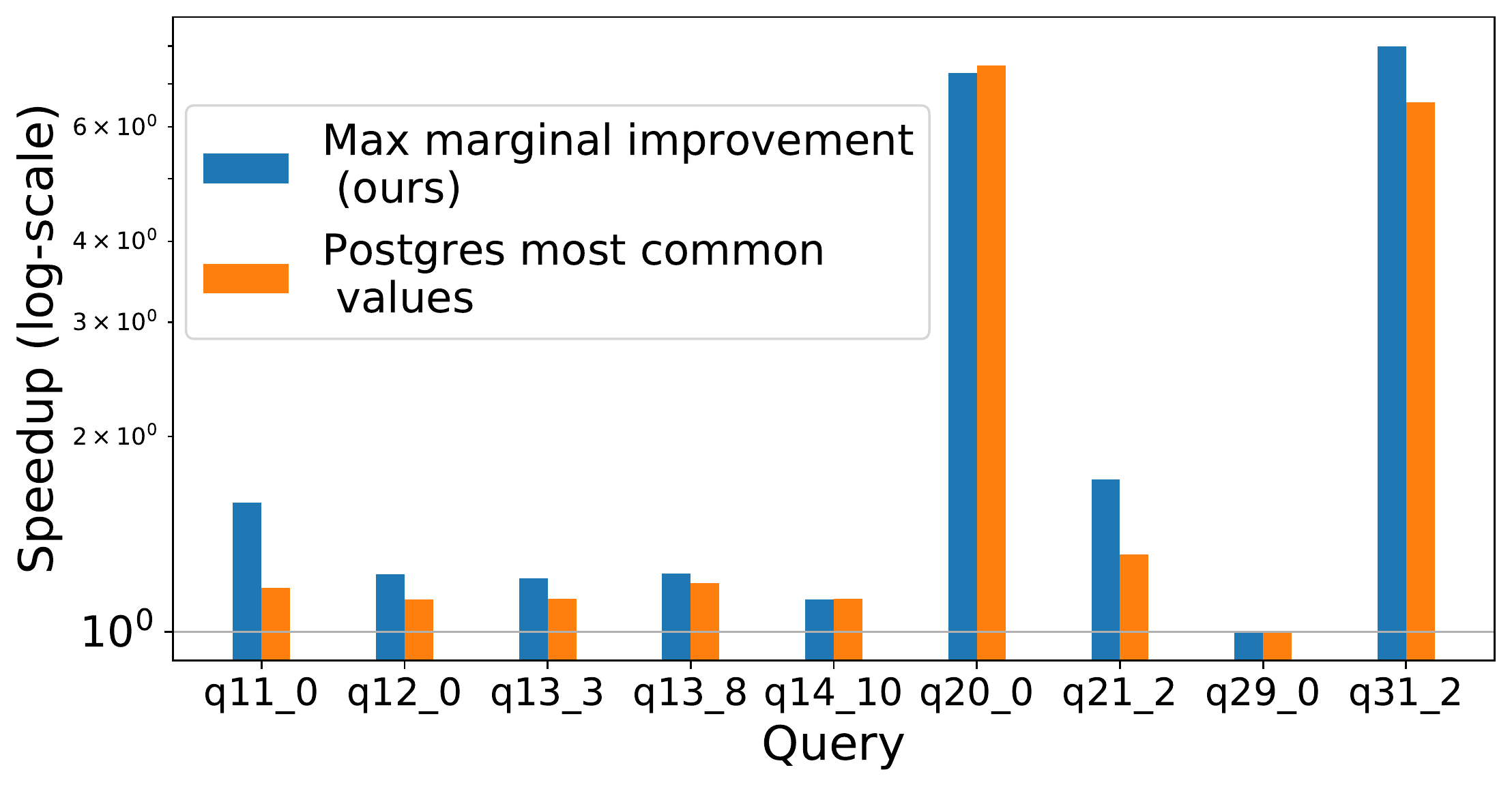}
        \caption{Embedding vocabulary.}
        \label{fig:vocabulary}
    \end{subfigure}    
    \caption{Speedup ablations on features and vocabulary.}
    \label{fig:ablations}
\end{figure}

In Figure~\ref{fig:selectivity_baseline}, we \rewrite{ablate our feature choice using raw features and selectivity features.} We compare their model speedups to $\sopt$ on a subset of Stack queries. Selectivity features perform far worse due to the poor cardinality estimates in PostgreSQL.

\paragraph{Vocabulary.} 
In Figure~\ref{fig:vocabulary}, \rewrite{we ablate how we select the embedding vocabulary for string features}. In particular, we evaluate our max marginal improvement method (described in Section~\ref{sec:modeling}) against choosing the most frequent values based on the PostgreSQL histogram. Our results confirm that the best strategy is choosing the vocabulary to the be the values with the most potential impact on the speedup.

\paragraph{How much training data is required?}
\begin{figure}[!t]
    \centering
    \includegraphics[width=0.8\linewidth]{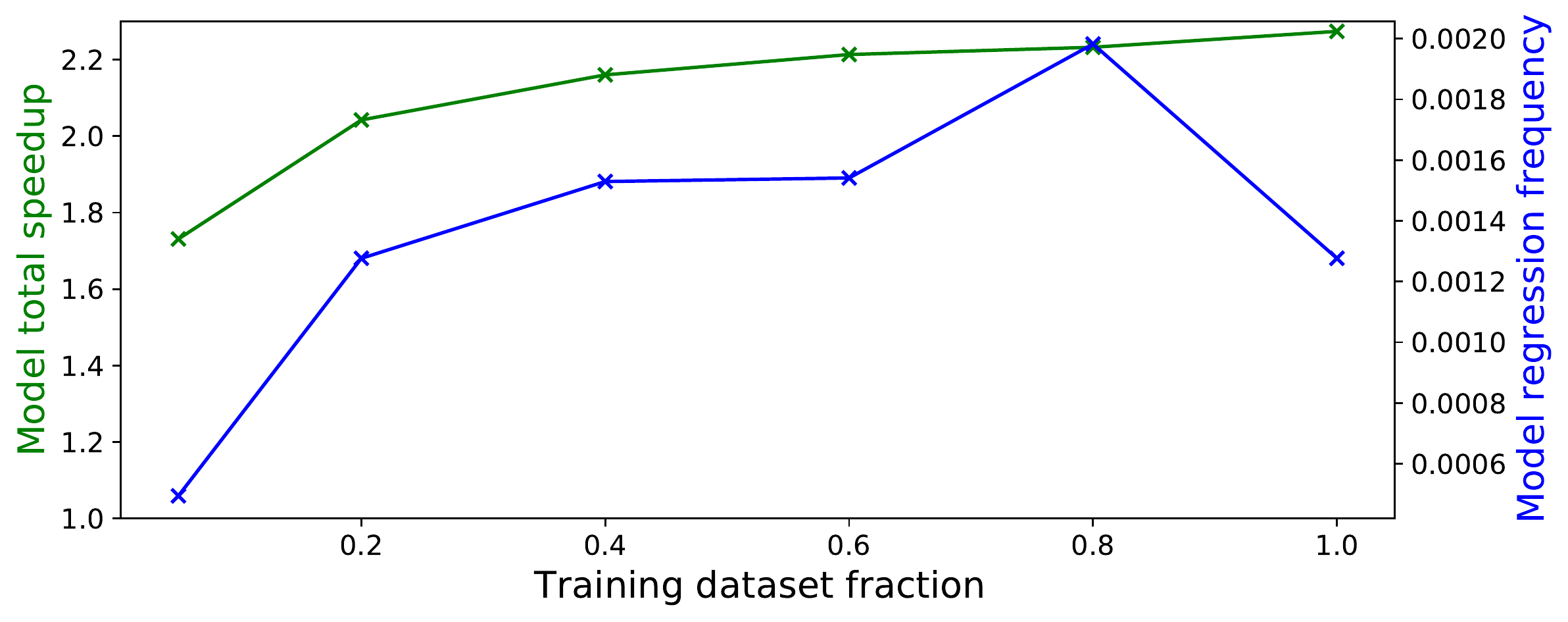}
    \caption{Model speedup and regression frequency as a function of fraction of training data used.}
    \label{fig:training_data_size}
\end{figure}

We evaluate the performance of our models when using less data by subsampling the training data size, as shown in Figure~\ref{fig:training_data_size}. As expected, model speedup improves with more training data while maintaining a regression frequency below 0.2\%. The leftmost point uses only 5\% of the data, or 200 training query instances, demonstrating that a large amount of speedup can be robustly captured with small amounts of training data.

\subsection{Dataset Contribution} 
\label{subsec:dataset_contribution}

\rewrite{To train models and evaluate \sysname, we generated a query execution dataset that comprises $\sim$14.2 years of CPU time across 200 million query executions of 131 query templates. By releasing this dataset online\footnote{https://github.com/google/kepler}, we envision that it can be used to facilitate future research in modeling and efficiency without having to execute any queries. For example, possible use cases include simulating active learning approaches that only selectively execute a subset of queries, or developing better models or loss functions. We believe that the techniques developed using this dataset -- and not the specifically trained models -- will be directly transferable to other parameterized query settings.}


%% file: sections/8_conclusion.tex
\section{Conclusion and Future Work} 
\label{sec:future_work}

We introduced \sysname, a system that can robustly speed up parameterized queries using a learning-based approach. We extensively evaluated \sysname on PostgreSQL and demonstrated that (1) our novel candidate generation algorithm RCE can provide significant speedups in query execution latency, and (2) robust ML models can reliably predict faster plans while avoiding regressions. \rewrite{Interestingly, we observed that RCE-generated plans were often far better than exact cardinality plans, indicating that even a widely-used system as PostgreSQL has significant room for improvement. Evaluating \sysname on database platforms other than PostgreSQL is a natural next step; we believe that the empirical nature of \sysname allows it to discover performance gains regardless of the DBMS.}

\rewrite{There are a myriad of future directions for improving the efficiency and performance of \sysname. For example, prior work in cardinality estimation can benefit \sysname in several ways. Instead of perturbing uniformly, RCE can leverage generative cardinality distributions to sample higher likelihood perturbations, e.g. from NeuroCard \cite{yang2020neurocard}. Another possibility is to augment the model features with the query plan tree and selectivity estimates, allowing the model to determine when cardinality estimates are accurate, as well as leveraging shared structure between similar query templates. Our models are the first demonstration that speedups can be robustly captured; they can likely be substantially improved via additional modeling techniques and tuning. Similarly, while our end-to-end PostgreSQL integration is sufficient to demonstrate \sysname's performance gains on a real system, every aspect of this implementation can be further tuned.}

\rewrite{We utilized an expensive training data collection procedure in order to make more robust claims about our results and produce a complete dataset for further modeling and efficiency research. For practical purposes, our training procedure can likely be made much more efficient, e.g. via active learning. In conjunction with Figure~\ref{fig:training_data_size}, this implies that similar performance can be achieved with significantly less training cost.}